\documentclass[12pt]{iopart}
\usepackage{cite}
\usepackage{color}

\usepackage{amssymb}
\usepackage{graphicx}

\def\SEC#1#2{\section{#1} \label{#2}}
\def\SUB#1#2{\subsection{#1} \label{#2}}

\def\NAME#1#2{\caption{#2} \label{#1}}

\def\FIGT#1{\begin{figure}[!t] \begin{center} #1 \end{center} \end{figure}}

\def\EPSW#1#2#3{\includegraphics[width= #2 cm,clip]{#1-eps-converted-to.pdf} \NAME{#1}{#3}}

\def\EN#1{\begin{enumerate} #1 \end{enumerate}}

\def\EQ#1{\begin{equation} #1 \end{equation}}
\def\EQL#1#2{\begin{equation} \label{#1} #2 \end{equation}}

\def\AR#1{ \begin{array}{l} #1 \end{array} }
\def\ARM#1#2{ \begin{array}{#1} #2 \end{array} } 

\def\LL{\left[}
\def\LR{\right]}
\def\ML{\left \{}
\def\MR{\right \}}

\def\LP#1{\left \{ \AR{#1} \right.}
\def\LPM#1#2{\left \{ \ARM{#1}{#2} \right.} 

\usepackage{graphicx}
\usepackage[hang,small,bf]{caption}

\usepackage[subrefformat=parens]{subcaption}

\usepackage{cases}
\begin{document}

\title{
Body-rotation behavior of pedestrians for collision avoidance in passing and cross flow
}

\author{Hiroki Yamamoto$^1$, Daichi Yanagisawa$^{2,1,*}$, Claudio Feliciani$^{2}$, and Katsuhiro~Nishinari$^{2,1}$}

\address{$^1$ Department of Aeronautics and Astronautics, School of Engineering,\\The University of Tokyo, 7-3-1 Hongo, Bunkyo-ku, Tokyo 113-8656, Japan}
\address{$^2$ Research Center for Advanced Science and Technology,\\The University of Tokyo, 4-6-1 Komaba, Meguro-ku, Tokyo 153-8904, Japan}

\ead{$^*$tDaichi@mail.ecc.u-tokyo.ac.jp}

\vspace{10pt}
\begin{indented}
\item[]February 2019
\end{indented}

\begin{abstract}
This study investigated the body-rotation behavior adopted by pedestrians to avoid others while moving in congested areas.
In such scenarios, body orientation often differs from walking direction, e.g., a pedestrian may step sideways.

The deviation between body orientation and walking direction during collision avoidance was studied by quantitatively evaluating the body rotation for counter-flows in narrow corridors.
Simple experiments, in which two pedestrians passed each other, were conducted.
It was found that pedestrians rotated their bodies when the corridor width was smaller than the sum of the widths of the two pedestrians.
This behavior was explained by analyzing the geometry of two ellipses circumscribing each other in a narrow corridor.
A preliminary model was developed, and the deviation between the body orientation and walking direction during passing was successfully simulated.

Finally, a cross-flow experiment, which is much more complex and realistic than the passing experiments, was performed; it was confirmed that body rotation behavior is also a critical factor in complex and realistic scenarios.
\end{abstract}

%

\SEC{Introduction}{INTRO}

Pedestrian dynamics, which is considered as one of the practical topics of active matters \cite{Schweitzer2007}, has been investigated extensively over the last two decades \cite{Helbing2001, Schadschneider2010} by means of simulations and experiments.

Simple models for bidirectional flow and evacuation were developed in the early stages of research on this topic.
The social force \cite{Helbing1995, Helbing2000} and floor field \cite{Burstedde2001} models are representative of force-based models in continuous space and cellular automata models, respectively.
They are useful for simulating collective real-world pedestrian phenomena, such as lane formation in bidirectional flow and the arching phenomenon that occurs at a bottleneck.

Behavioral experiments have also been conducted with real pedestrians to elucidate pedestrian behaviors in crowds and calibrate various model parameters.
The results of various circuit experiments have allowed researchers to produce refined fundamental diagrams, showing a relation between density and velocity (flow) \cite{Seyfried2005, Jelic2012a}.
Lane formation was observed in experiments on bidirectional flow \cite{Hoogendoorn2005, Kretz2006a, Zhang2012a, Feliciani2016pre,Feliciani2018plosone}.
A bidirectional-flow experiment with an oblique intersecting angle was also conducted \cite{Wong2010} and investigated with macroscopic models \cite{Hanseler2017}.
Furthermore, the zipper effect (alternate merging of two lanes into one lane) and the relation between the width of a bottleneck and flow were revealed by bottleneck-flow experiments \cite{Hoogendoorn2005TransSci, Kretz2006b, Seyfried2009}.

Recently, more detailed pedestrian characteristics have been considered and experimentally investigated to develop more realistic models.
The social force model was extended by introducing a required space for walking \cite{Seyfried2006} and a self-stopping mechanism \cite{Parisi2009}, resulting in the successful reproduction of realistic fundamental diagrams.  
On the other hand, many elements,
e.g., collisions between pedestrians \cite{Kirchner2003a, Kirchner2003b, Yanagisawa2009}, forces between pedestrians \cite{Henein2007,Henein2010}, advance avoidance by anticipation \cite{Suma2012, Nowak2012}, heterogeneity of walking speed \cite{Weng2006, Jiang2007}, and aggressiveness in a situation of conflict \cite{Hrabak2016}, have been implemented in the floor field model.
Moreover, the agent-based approach \cite{Bandini2009} has allowed researchers to represent the heterogeneous properties of pedestrians, e.g., age \cite{Bandini2015}.
Furthermore, the behaviors of individual pedestrians have been researched in detail to model pedestrians as beyond simple self-driven particles.
The effects of stride (step length) and walking tempo have been studied \cite{Yanagisawa2012, Jelic2012b} and introduced into various models \cite{Sivers2015}.
The decision-making process for pedestrians has been investigated through experiments on the inflow process \cite{Liu2016b, Ezaki2016cd}, which is a counter process of evacuation \cite{Ezaki2012a, Ezaki2015tgf}.





Although many pedestrian characteristics have been studied and implemented in various models,
research on body-rotation behaviors has only been started very recently \cite{Jin2017}
\footnote{
Jin \textit{et al.} observed body rotations in their bidirectional-flow experiment and modeled them with a cellular automaton model.
}
, and further detailed investigations are required.
Here, it should be emphasized that the body rotation, which is the focus of this study, is not the change in the walking direction due to turning but the change in the body orientation along the same walking direction.

The effect of body rotation is especially important in congested situations.
Bidirectional-flow simulations often end up with a deadlock, where no pedestrians can move without additional rules.
In cellular automata models, two opposing pedestrians stochastically exchange their positions to resolve a deadlock \cite{Blue2001, Feliciani2016PhysicaA, Yanagisawa2016cd}.
Although no detailed explanations exist on how real pedestrians exchange positions, it can be assumed that position exchange is formed by penetrating into the opposing crowd with body rotation.
Even when pedestrian density is not extremely high, pedestrians often employ body rotation to avoid colliding with one another without greatly changing their walking direction.
This implies that the degree and frequency of body rotation may be related to the pedestrian density.
Assuming this to be true, it should be possible to approximate the density with body-rotation data recorded by personal smartphones equipped with gyro sensors \cite{Nagao2018pa}.

Many models represent pedestrians using circles as they are easy to simulate due to their rotational symmetry.
However, this symmetry precludes representing the effect attributed to pedestrian orientation.
In other words, circular pedestrians always occupy the same region irrespective of their orientation, 
which makes it impossible to study the effect of body rotation.

One solution to this problem is ``elliptic pedestrians'', i.e., pedestrians represented using ellipses.
Ellipses allow the difference between the effective shoulder width and effective bust depth during walking to be represented.
The use of elliptic pedestrians has been proposed in previous works~\cite{Fruin1971, Pauls2004, Was2006, Chraibi2010}.
For example, Chraibi \textit{et al.}~\cite{Chraibi2010} utilized elliptic pedestrians in their force-based model and succeeded in reproducing the fundamental diagram of unidirectional flow, which quantitatively agrees well with  the experimental data.
Herein, the pedestrians were represented with ellipses to allow geometrical differences to be considered;
it was confirmed that realistic body rotation was reproduced with elliptic pedestrians.



The focus of this study was originally restricted to a simple case, i.e., collision avoidance between two pedestrians when passing through a narrow corridor.
This was considered to be one of the most fundamental situations appropriate for research on body-rotation behavior, since the minimal interaction is collision avoidance between two pedestrians.
In-detail investigations of microscopic interactions would contribute to the understanding of macroscopic phenomena.
Moussaid \textit{et al.} \cite{Moussaid2009} thoroughly analyzed collision avoidance between two pedestrians when changing walking direction and speed to discuss collective patterns in bidirectional crowd flows. 

Although the walking behavior of a single pedestrian has been investigated thoroughly in various studies \cite{Imai2001}, such studies have not focused on the interactions between pedestrians.
There are also many sophisticated works on collision avoidance between two pedestrians;
however, the motivations of those works were different from those of ours.

One motivation of these previous works was the investigation of personal space.
Gorrini \textit{et al.} \cite{Gorrini2014b} performed experiments where two pedestrians approached each other and clarified that the size of the front zone of personal space changes depending on the pedestrians' walking speeds.

The other motivation was studying collision avoidance in a wide area.
Olivier \textit{et al.} \cite{Olivier2013} performed crossing experiments at an angle of $90^\circ$ and revealed that the latter pedestrian gave way to the former pedestrian by decreasing his/her walking speed.
Daamen \textit{et al.} \cite{Daamen2014} and Huber \textit{et al.} \cite{Huber2014} performed crossing experiments at various crossing angles.
The former study determined that pedestrians tended to pass each other on the right-hand side when the crossing angle increased.
The latter study showed that pedestrians adjusted both the walking direction and speed for acute crossing angles, but only adjusted the walking direction for obtuse crossing angles.
Parisi \textit{et al.} \cite{Parisi2016} conducted experiments with groups of pedestrians and found abrupt evasive maneuvers to increase with the number of pedestrians.
They also revealed that pedestrians preferred to change their walking direction rather than decrease their speed.
Furthermore, Zanlungo \textit{et al.} \cite{Zanlungo2012} extended the model in \cite{Moussaid2009} by introducing preferred direction in collision avoidance and overtaking, they then compared the results of their simulation with real-world data.
Since wide areas were assumed in the above studies, pedestrians were able to avoid conflicts by changing their walking direction and speed without body rotation.

In contrast to these previous studies, this study investigated collision avoidance between two pedestrians in a narrow corridor, where collision avoidance would fail without body rotation.
The results of this study may contribute to a better understanding of pedestrian behavior and be a source of inspiration to improve existing models.
We would also like to mention that the participants of our experiment included both males and females with various ages, thus, our results are not limited to a specific type of pedestrians.

A cross-flow experiment was conducted to examine the importance of body-rotation behavior in more complex and realistic scenarios.
We succeeded to observe body rotations in the experiment and to reveal a condition of body rotation in cross flow.

The remainder of this paper is organized as follows.
Section \ref{MODEL} describes the analysis of the geometric condition in which two elliptic pedestrians circumscribe each other while passing through a narrow corridor.
Section \ref{EXP} explains the experimental setup and conditions.
Section \ref{COMP} elaborates upon the analysis of the rotational angles in the experiment from various perspectives and examines the validity of the elliptic representation of pedestrians.
Section \ref{Deceleration} expounds upon the analysis of the deceleration due to body rotation during passing.
Section \ref{SIM} discusses the developed model and how it simulates passing in a narrow corridor.
Sections \ref{EXP1-2} and \ref{COMP1-2} describes an experiment and analyses, similar to those described in Secs. 3 and 4, conducted for various types of participants.
Section \ref{CROSSFLOW} presents the results of the cross-flow experiment.
Finally, Sec. \ref{CONC} provides concluding remarks and discussions.
\SEC{Geometric analysis of collision avoidance between elliptic pedestrians with body rotation}{MODEL}

\SUB{Elliptic excluded volume of pedestrians}{Ellipse}

\begin{figure}[t]
\begin{tabular}{cc}
\begin{minipage}[t]{0.5\linewidth}
\centering
\includegraphics[keepaspectratio, scale=0.6]{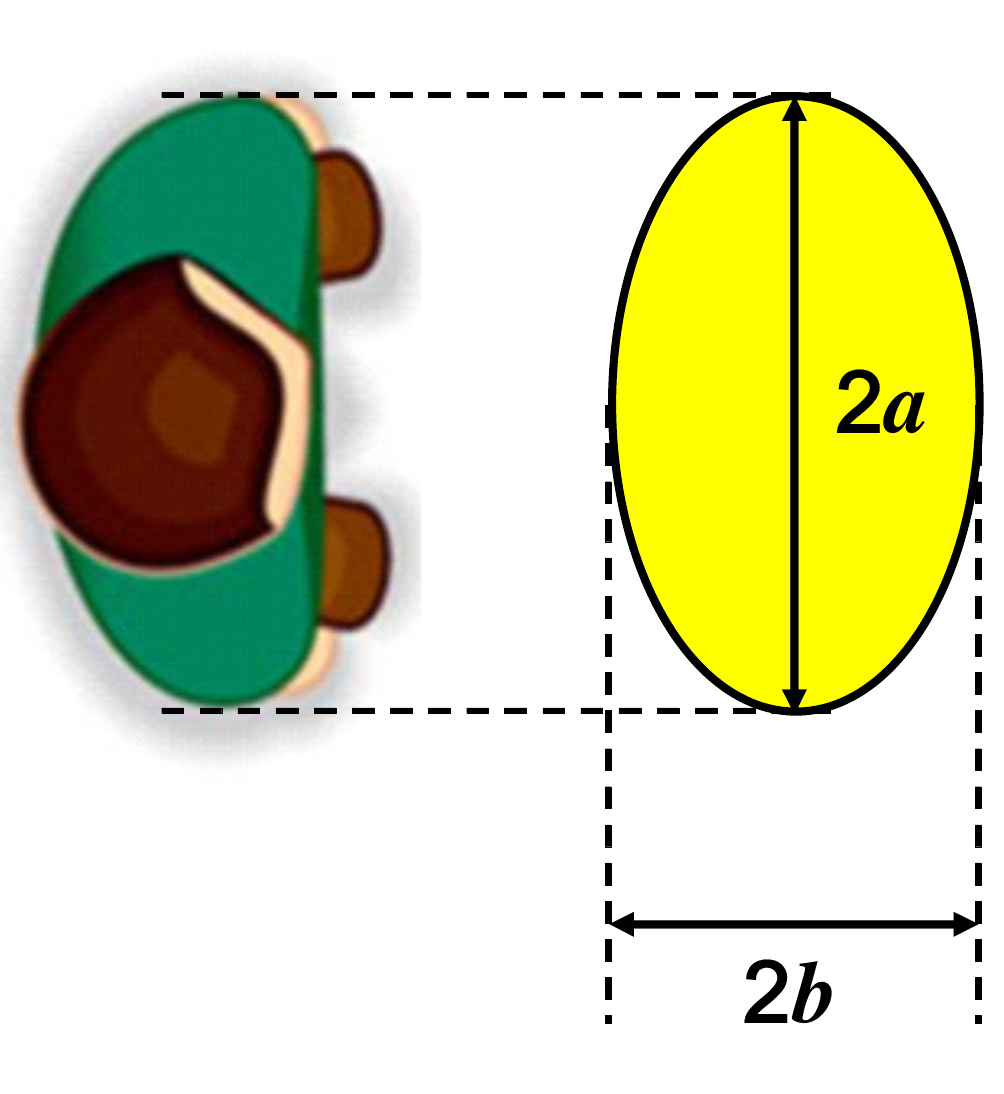}
\subcaption{}\label{fig:figure1a}
\end{minipage} &
\begin{minipage}[t]{0.5\linewidth}
\centering
\includegraphics[keepaspectratio, scale=0.6]{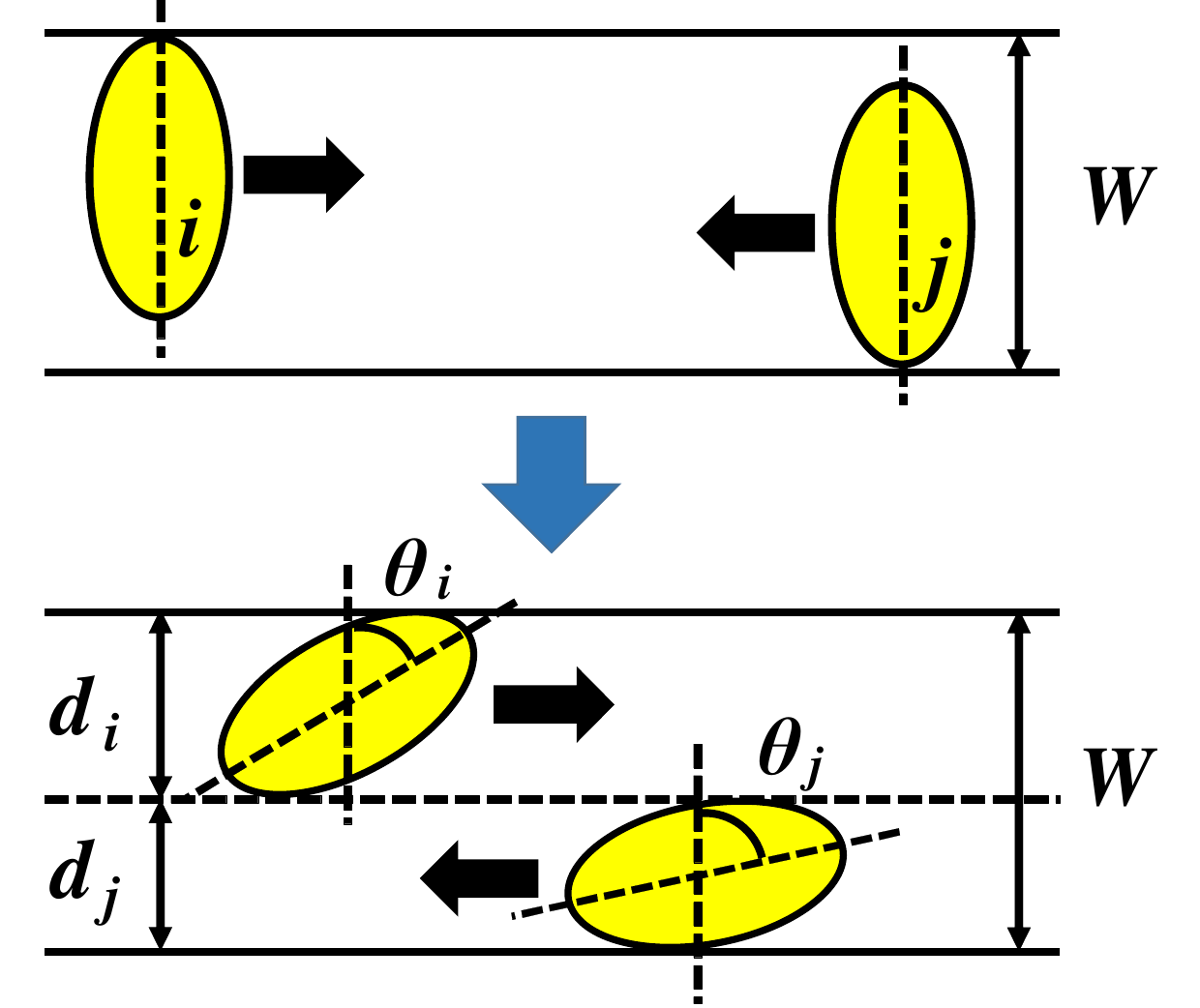}
\subcaption{}\label{fig:figure1b}
\end{minipage}
\end{tabular}
\caption{
(a) Schematic of an elliptic pedestrian.
The constant parameters $a$ and $b$ $(a > b)$ represent the semi-major and semi-minor axes of the ellipse, respectively. 
(b) Schematic of collision avoidance in passage through a narrow corridor.
The top and bottom figures show the scenarios before and during passing, respectively.
The parameter $W$ is defined as the corridor width.
There are two elliptic pedestrians $i$ and $j$; their passing rotational angles, i.e., the slope angles of the major axes from the initial directions, are represented  as $\theta_i$ and $\theta_j \in [0^\circ, 90^\circ]$, respectively;
their rotated widths are described by $d_i \in [2b_i, 2a_i]$ and $d_j \in [2b_j, 2a_j]$, respectively.
}
\end{figure}

Elliptic pedestrians were considered, as shown in Fig. \ref{fig:figure1a}, and collision avoidance with body rotation in a narrow corridor was investigated, as shown in Fig. \ref{fig:figure1b}.
The semi-major and semi-minor axes of an ellipse are described by parameters $a\ (>0)$ and $b\ (>0)$, such that $2a$ and $2b$ correspond to a pedestrian's effective shoulder width and bust depth, respectively. 
$a > b$ was assumed, so that the width of a pedestrian decreases with rotation.
Note that W{\c{a}}s \textit{et al.} also modeled the physical excluded volume of a pedestrian using an ellipse \cite{Was2006}.

It has been reported that the effective excluded volume of a walking pedestrian varies with velocity~\cite{Pauls2004}; the velocity-dependent radii of walking pedestrians in a one-dimensional circuit has been considered in \cite{Seyfried2006}.
Moreover, Chraibi \textit{et al.} modeled a pedestrian using an ellipse with velocity-dependent axes for a two-dimensional case\cite{Chraibi2010}.
Given that the dependency of the axes on velocity is different along the longitudinal (walking direction) and lateral (direction vertical to the walking direction) axes, the semi-major and semi-minor axes were exchanged in their model.
In other words, the lateral axis is the semi-major axis when the velocity is small, while the longitudinal axis is the semi-major axis when the velocity is large.

One can model a pedestrian more realistically by introducing velocity-dependent axes; however, this study adopted constant axes for simplicity, since the velocity is low in passage through a narrow corridor, as shown in Fig. \ref{fig:figure1b};
when the velocity is low, the velocity-dependent effect considered in \cite{Seyfried2006, Chraibi2010} would also diminish.
Furthermore, the variation in the lateral axis
 is mainly attributed to the swaying motion of pedestrians, according to \cite{Chraibi2010, Hoogendoorn2005TransSci}. 
It can be assumed that pedestrians do not sway in the lateral direction when passing through a narrow
corridor, and thus the lateral axis was assumed to be constant in this study.

\SUB{Relation between passing rotational angle and corridor width}{Rotation}

This subsection explains how the relationship between the rotational angle of a body $\theta$ and corridor width $W$ was obtained.

First, let's focus on the width $d$ occupied by a single elliptic pedestrian when rotating by $\theta$.
By analyzing the geometry of an ellipse, the relation between width $d$ and passing rotational angle $\theta$ was calculated: 
\EQL{eq:d}{
d(\theta, a, b) = 2 \sqrt{a^2\cos^2\theta+b^2\sin^2\theta}.
}
The detailed derivation is described in \ref{MATH}.
Taking the derivative of $d$ with respect to $\theta$:
\EQ{
d'(\theta, a, b)=\frac{(b^2-a^2)\sin2\theta}{ \sqrt{a^2\cos^2\theta+b^2\sin^2\theta} }.
\label{eq:ddash}
}

The width $d$ and its derivative $d'$ are plotted as functions of $\theta$ in Fig. \ref{figure2}a and \ref{figure2}b, respectively.
The parameters were set to $(a,b)=$(24.9 cm, 15.5 cm), as determined by the empirical data obtained from the experiment explained in Secs. \ref{EXP} and \ref{COMP}.
It can be seen that the slope of $d$ is always negative; however, the steepness varies with the passing rotational angle $\theta$ (Fig. \ref{figure2}a);
Fig. \ref{figure2}b illustrates this feature more clearly.  
This result indicates that $d$ is sensitive (insensitive) to the change of $\theta$ where $\theta$ is around $45^\circ$ ($0^\circ$ and $90^\circ$). 
Therefore, the width of an elliptic pedestrian does not significantly change with oscillatory rotation during normal walking $(\theta \approx 0)$.
Note that we use the units cm and deg ($^\circ$) for lengths and angles throughout this paper.

\FIGT{\EPSW{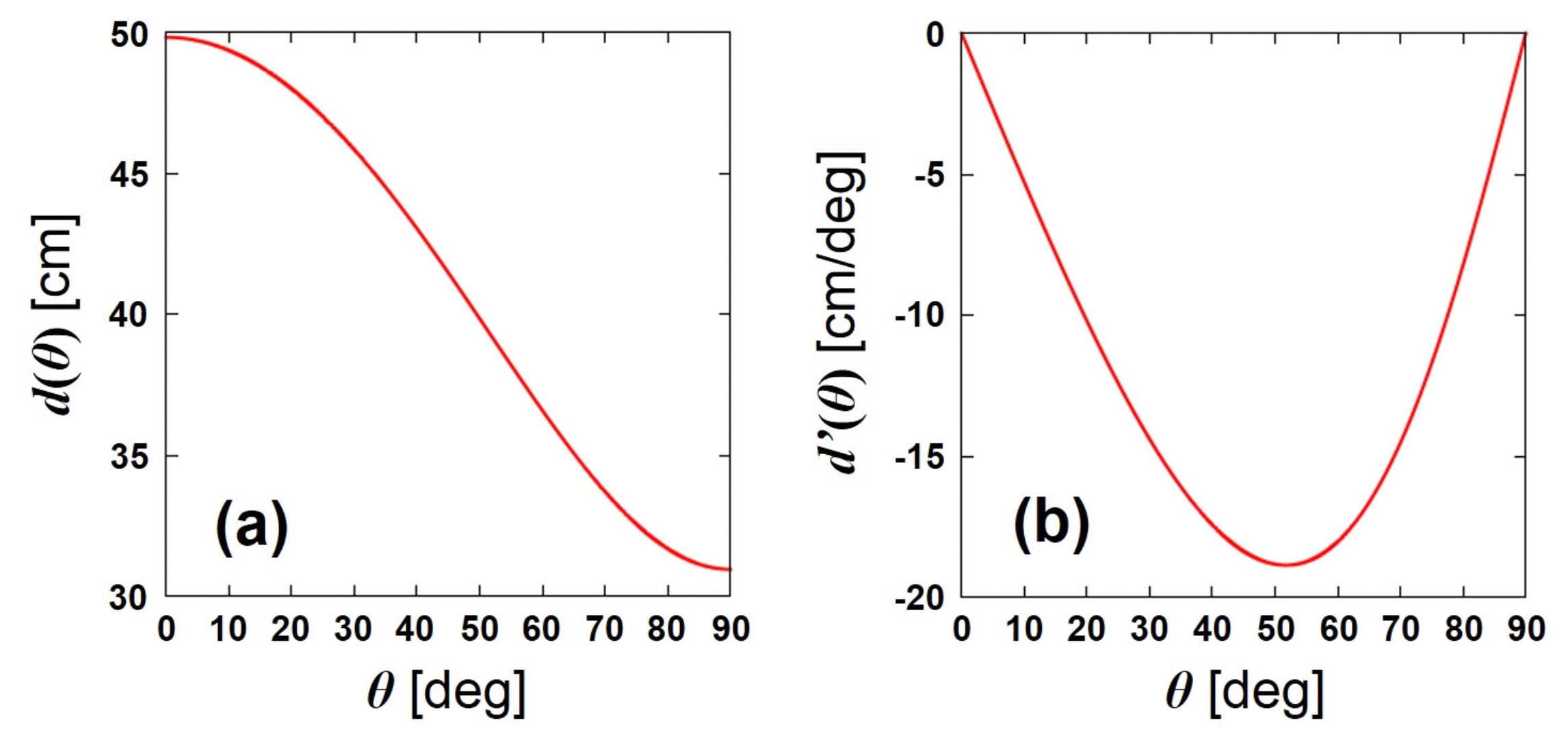}{16}{
(a) Width of a rotated ellipse $d$ as a function of $\theta$.
(b) Derivative of the width of a rotated ellipse $d'$ as a function of $\theta$.
The parameters in both figures are set as $(a,b)=$(24.9 cm, 15.5 cm), which were determined by the experimental data in Sec. \ref{Comparison}.
}}

Next, let's deal with two elliptic pedestrians ($i$ and $j$) who rotate during passing to avoid each other.
The semi-major and semi-minor axes of pedestrian $i$ $(j)$ is $a_i$ $(a_j)$ and $b_i$ $(b_j)$, respectively.
It was assumed that these axes rotate by $\theta_i$ and $\theta_j$, respectively, such that the sum of their widths is equal to the corridor width $W$ during passing, as in Fig. \ref{fig:figure1b}.
More specifically, it was assumed that the two ellipses circumscribed each other as follows:
\EQL{eq:W1}{
d_i(\theta_i, a_i, b_i) +d_j(\theta_j, a_j, b_j) = W.
}

Here, the focus was on the situation in which the inequality $2(b_i + b_j) \le W \le 2(a_i + a_j)$ is satisfied; otherwise, rotation does not work effectively.
If $W < 2(b_i + b_j)$, it is impossible for two elliptic pedestrians to pass each other even with the maximum rotational angle of $90^\circ$ rotation.
In contrast, if $W > 2(a_i + a_j)$, two elliptic pedestrians do not need to rotate to pass each other, i.e., $\theta = 0^\circ$.

In Secs. \ref{Comparison} and \ref{Comparison1-2}, the analytic and experimental results of this study are compared, and the validity of using elliptic pedestrians with (\ref{eq:W1}) for modeling body rotation is examined.

\begin{figure}[b!]
\begin{tabular}{cc}
\begin{minipage}[t]{0.5\linewidth}
\centering
\includegraphics[keepaspectratio, height=5cm]{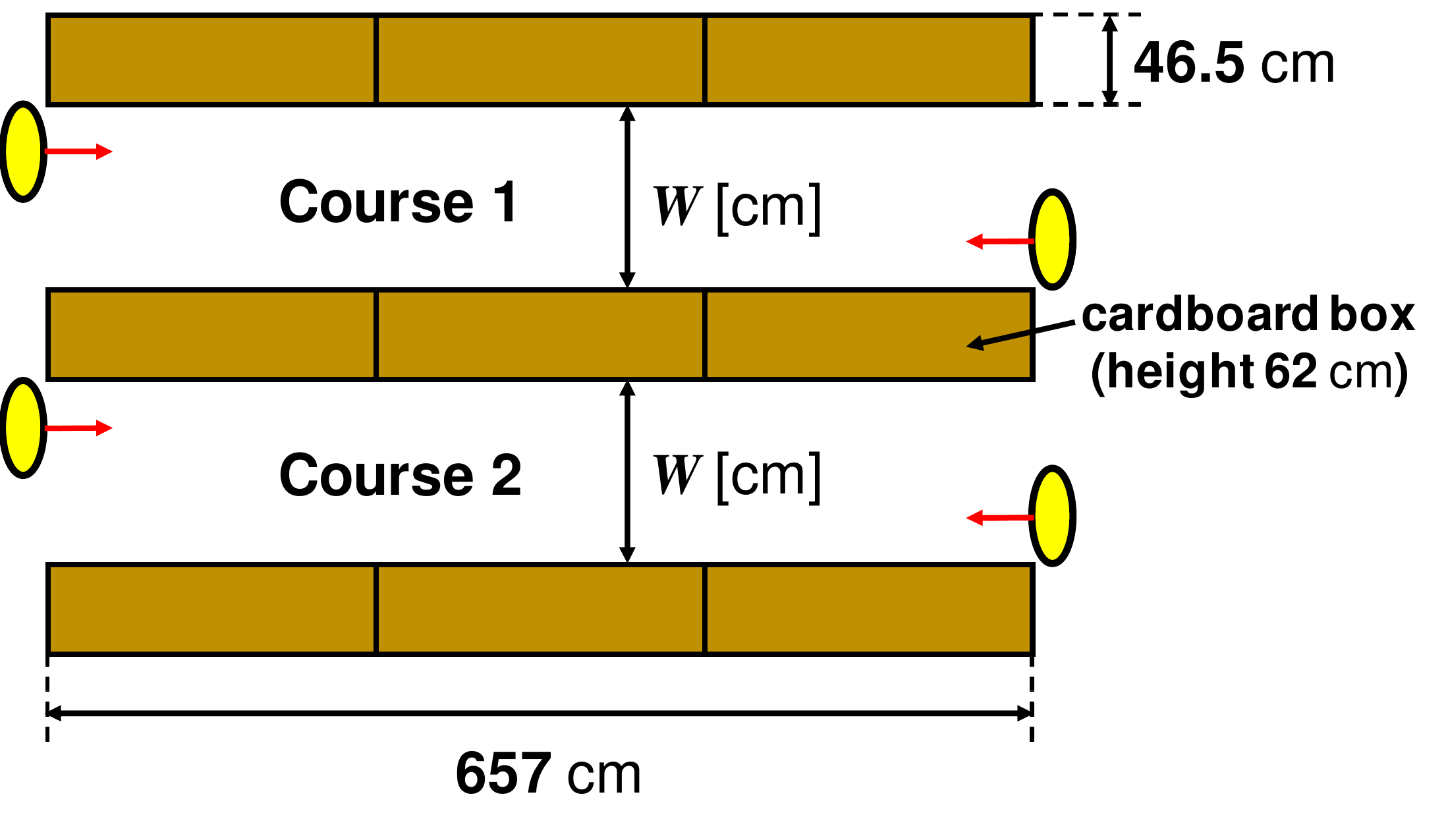}
\subcaption{}\label{fig:figure3a}
\end{minipage} &
\begin{minipage}[t]{0.5\linewidth}
\centering
\raisebox{7mm}{\includegraphics[keepaspectratio, height=4cm]{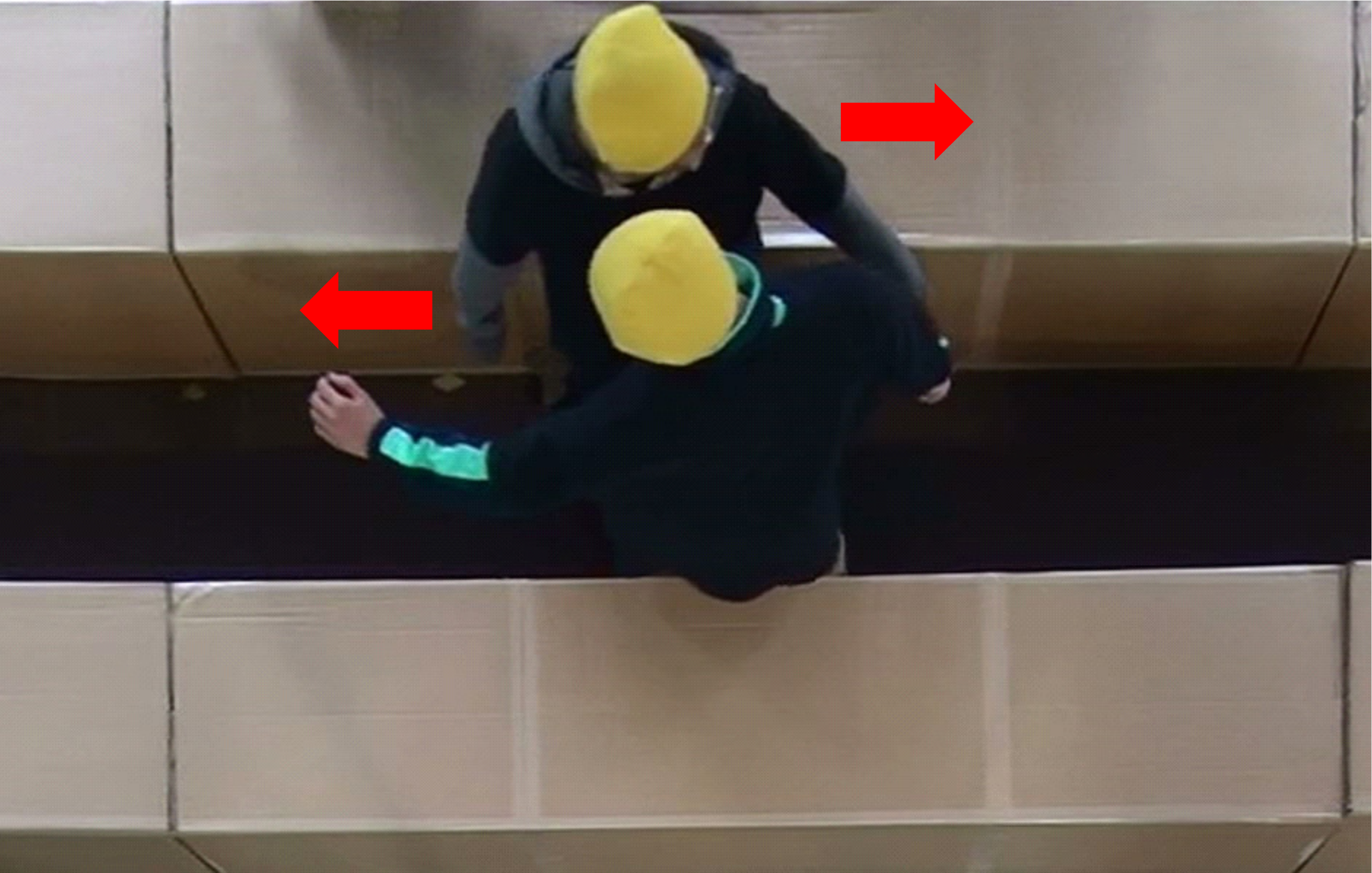}}
\subcaption{}\label{fig:figure3b}
\end{minipage}
\end{tabular}
\caption{
(a) Schematic of the experimental setup.
Two courses were formed by placing cardboard boxes.
Four participants (yellow ellipses) standing at the ends of the courses walked to the other ends upon cue.
The passing side was specified to the participants in advance to each trial.
In this figure, the starting positions in the left-passing case are depicted.
(b) Snapshot of the experiment for $W$ = 60 cm in the left-passing case.
It can be observed that the two participants pass each other by rotating their bodies.
}
\end{figure}

\SEC{
Setup and conditions of corridor experiment 1
}{EXP}
A real experiment was performed to investigate the body rotation and side-stepping behavior of
pedestrians from the perspective of collision avoidance in a narrow corridor.
The experiment was conducted at the Research Center for Advanced Science and Technology (RCAST), The University of Tokyo, Japan.
Two courses were constructed using cardboard boxes, as described in Fig. \ref{fig:figure3a}.

The experiment included four male participants (ID number = 1, 2, 3, and 4), ages ranging from 18 to 25.
Initially, the participants were arranged at the end of each course.
One trial of the experiment started when the participants were instructed to go.
The participants first started to walk, passed each other midway through the course, and exited the course from the other end.
Figure \ref{fig:figure3b} shows a snapshot taken during the experiment. 
The participants were instructed to stand and walk on the left or right side of the course before each trial so that they could pass each other without dithering.

A commercial tablet (Nexus 7 (2013)) equipped with a gyro sensor was strapped between the chest and belly of each participant to record angular velocity.
The tablets were strapped on the participants' bodies such that the recorded angular velocities would be identical to the angular velocities of their bodies' rotations.
The precision and accuracy of the gyro sensor in the tablets were verified in \cite{Feliciani2016ped,Feliciani2018tmb}, and both were concluded to be below 1\%.

The width $W$ of the two courses were varied as 60, 70, 80, 90, 100, 110, 120, and 140 cm, and the angular velocity and travel time of each participant were observed.
By changing the passing side and participant pairing, six types of trial sets were executed twice for each $W$, as summarized in Tab. \ref{tab:exp-cond}.
Two pairs of participants (four participants) were included in each trial.
Thus, 2 (pairs) $\times$ 2 (trials) $\times$ 6 (trial sets) = 24 datasets were obtained for each $W$ and $24 \times 8$ (widths) = 192 datasets in total.

\begin{table}[t!]
\centering
\caption{
Overview of the six types of trial sets for each corridor width $W$.
The numbers in the parentheses represent the ID numbers of the four participants.
}
\label{tab:exp-cond}
\begin{tabular}{cccc}
\hline
Trial No. & Passing side & Pair in Course 1 & Pair in Course 2 \\ \hline
1, 2 & left & (1, 2) & (3, 4) \\ \hline
3, 4 & right & (1, 2) & (3, 4) \\ \hline
5, 6 & left & (2, 3) & (1, 4) \\ \hline
7, 8 & right & (2, 3) & (1, 4) \\ \hline
9, 10 & left & (1, 3) & (2, 4) \\ \hline
11, 12 & right & (1, 3) & (2, 4) \\ \hline
\end{tabular}
\end{table}

\SEC{
Analysis of passing rotational angles in corridor experiment 1 
}{COMP}
\SUB{Time series of angular velocities and rotational angles}{TimeSeries}

\begin{figure}[t]
\centering
\includegraphics[width=10cm,clip]{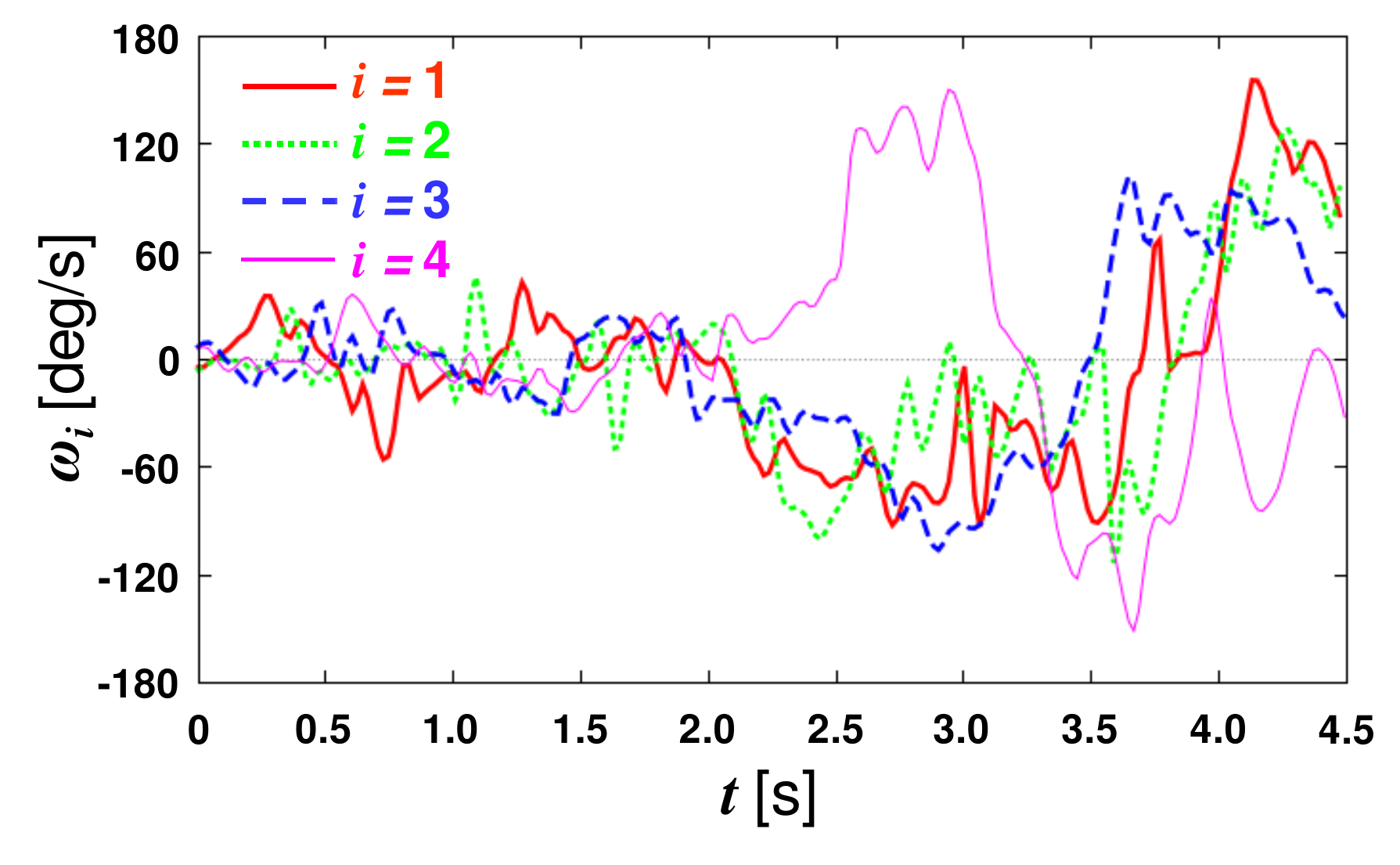}
\caption{
Time series of angular velocities $\omega_i(t)$ [deg/s] of the four participants obtained from the tablets in one trial for $W=60$ cm, Trial $\rm{No.}=$ 1 (left passing).
$\omega_i(t)>0$ ($\omega_i(t)<0$) corresponds to the counter-clockwise (clockwise) rotation of the participants.
}
\label{fig:figure4}
\end{figure}

In the experiment, the time series of angular velocities $\omega_i(t)$ ($i=1$, 2, 3, and 4 (ID number)) were obtained at approximately 50 Hz
from the tablets.
Figure \ref{fig:figure4} shows $\omega_1(t)$, $\omega_2(t)$, $\omega_3(t)$, and $\omega_4(t)$ in one trial with $W$ = 60 cm and left passing.
The positive and negative angular velocities correspond to the counter-clockwise and clockwise rotations, respectively.
In this example, participants 1 and 2 (3 and 4) were paired together.
It can be seen that the magnitudes of the angular velocities started to increase at around $t$ = 2.0 s;
this indicates the beginning of rotation.

In the left-passing case, the participants faced outward with respect to the course with positive (counter-clockwise) rotation and faced inward with respect to the course (the other participant) with negative (clockwise) rotation.
Thus, participants 1, 2, and 3 faced inward with respect to the course, while only participant 4 faced outward with respect to the course, as shown from the data corresponding to around $t$ = 2.5 s in Fig. \ref{fig:figure4}.
Note that the direction in which the participants faced during passing was maintained throughout the experiment.
Furthermore, no significant difference was found between the left-passing and right-passing cases.
Thus, we do not distinguish between the two cases in the following.

By numerically integrating the datasets of angular velocities $\omega_i(t)$, the time series of the rotational angle of each participant $\varphi_{i}(t) \in [-180^{\circ}, 180^{\circ}]$ was obtained as follows:  
\begin{equation}
\varphi_{i} (t) =  \int_{0}^{t} \omega_i (t) dt \approx \sum_n \omega_{i,n} \delta t_{i,n},
\label{eq:phi}
\end{equation}
where $\omega_{i,n}$ and $\delta t_{i,n}$ are participant $i$'s $n$-th angular velocity and time interval, respectively.

The time interval $\delta t_{i,n}$ should be small enough, i.e., the measurement frequency of the sensors should be large enough, to obtain accurate rotational angles.
It was confirmed that 50 Hz ($\delta t_{i,n} =$ 0.02 s) is large enough; however, 
due to a bug in the tablets' operating system,
the measurement frequency sometimes declined ($\delta t_{i,n}$ increased).
For such time series, it was impossible to compute the accurate rotational angles; therefore, the time series of the pair that included $\delta t_{i,n} \geq$ 0.1 s were discarded.
Eventually, two datasets were removed from $W=60$, 90, 140 cm, and one dataset was removed from $W=80$, 100 cm.

\begin{figure}[htbp]
\centering
\includegraphics[width=16cm]{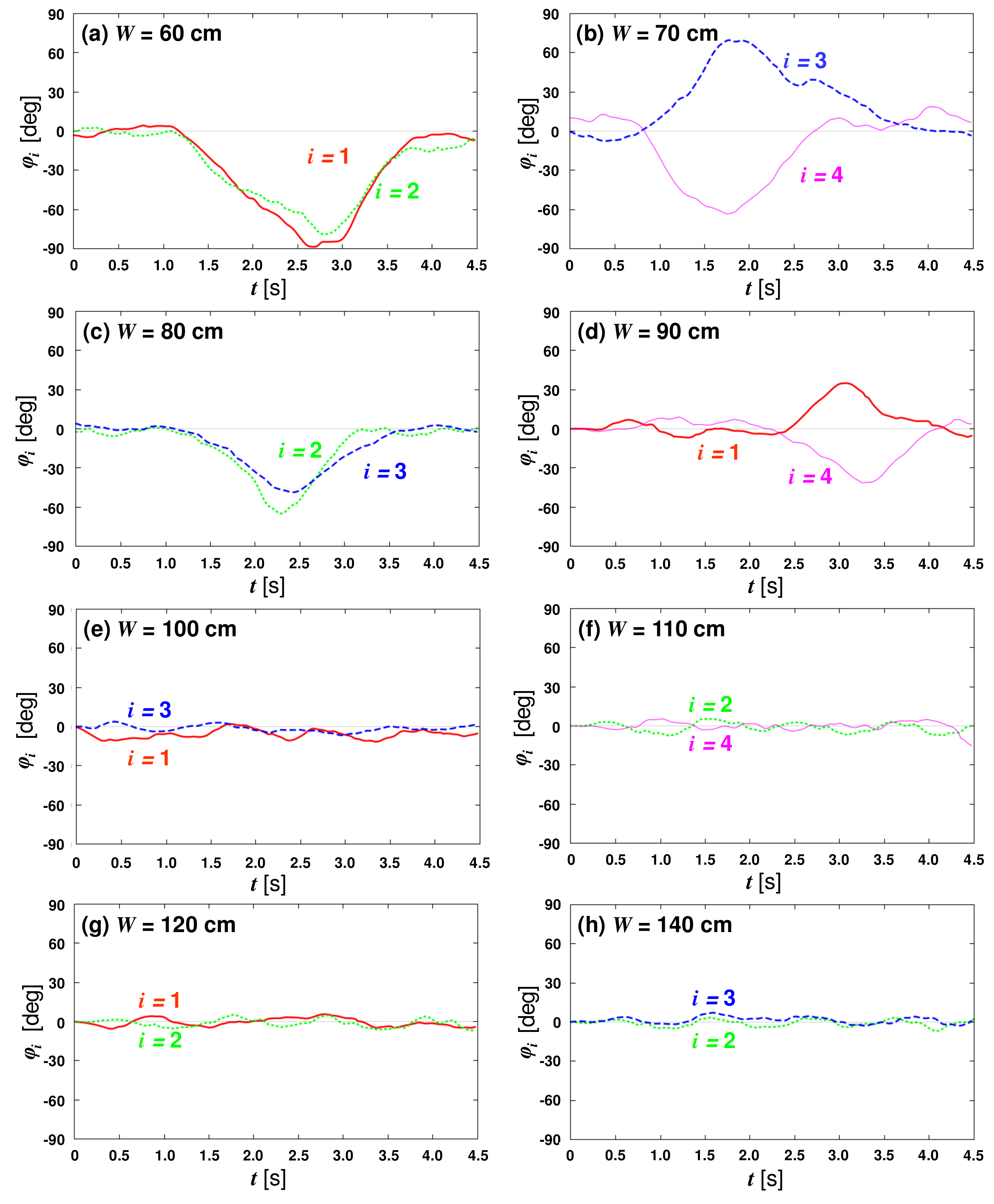}
\caption{
Time series of the rotational angle $\varphi_i(t)$ [deg] of one pair in one trial.
(a) $W=60$ cm, left-passing.
(b) $W=70$ cm, right-passing.
(c) $W=80$ cm, left-passing.
(d) $W=90$ cm, right-passing.
(e) $W=100$ cm, left-passing.
(f) $W=110$ cm, left-passing.
(g) $W=120$ cm, left-passing.
(h) $W=140$ cm, left-passing.
Note that $t=0$ in the figures does not always correspond to the beginnings of the trials because 4.5 s data was extracted around passing for visibility.
}
\label{fig:figure5}
\end{figure}

The $\varphi_{i}(t)$ of one pair in one trial was plotted for each $W$ as shown in Fig. \ref{fig:figure5}.
It can be seen that the participants clearly rotated their bodies when passing for $W=60$, 70, 80, and 90 cm, whereas they did not for $W=100$, 110, 120, and 140 cm.
Actually, a few data points indicated clear rotation for $W=100$ cm.
Thus, in the following, the focus is placed on $W=60$, 70, 80, 90, and 100 cm, where body rotation was observed during passing.
Note that deceleration also occurred from 100 cm.
See Sec. \ref{Deceleration} for details.



\SUB{Extraction of passing rotational angles}{PRAngles}
This subsection describes the derivation of the experimental passing rotational angles $\theta_i \in [0, 90^{\circ}]$
from the time series of the rotational angles $\varphi_i$.
To determine the passing rotational angle without ambiguity, it was assumed that the maximum absolute rotational angle $|\varphi|_{{\rm max}, i}$ was the passing rotational angle $\theta_i$, i.e., $\theta_i = |\varphi|_{{\rm max}, i}$.
Note that all passing rotational angles larger than 90$^\circ$ were replaced with 90$^\circ$.

However, the method has its limitations.
First, the oscillation of the rotational angle in normal walking is occasionally regarded as the passing rotational angle.
Second, in a few datasets, there were great discrepancies between the times at which each participant achieves the maximum rotational angle, making it inappropriate to regard them as sets of passing rotational angles.
Therefore, two thresholds were introduced, one for the passing rotational angle $\theta_i$ and another for the time gap between the points at which the two participants achieved the maximum rotational angles, which are denoted as $\Delta t _{|\varphi|_{\rm max}}$.
Any dataset that did not satisfy the conditions of the two thresholds were converted appropriately, as described in \ref{100cm}, before being used for analysis.

\FIGT{\EPSW{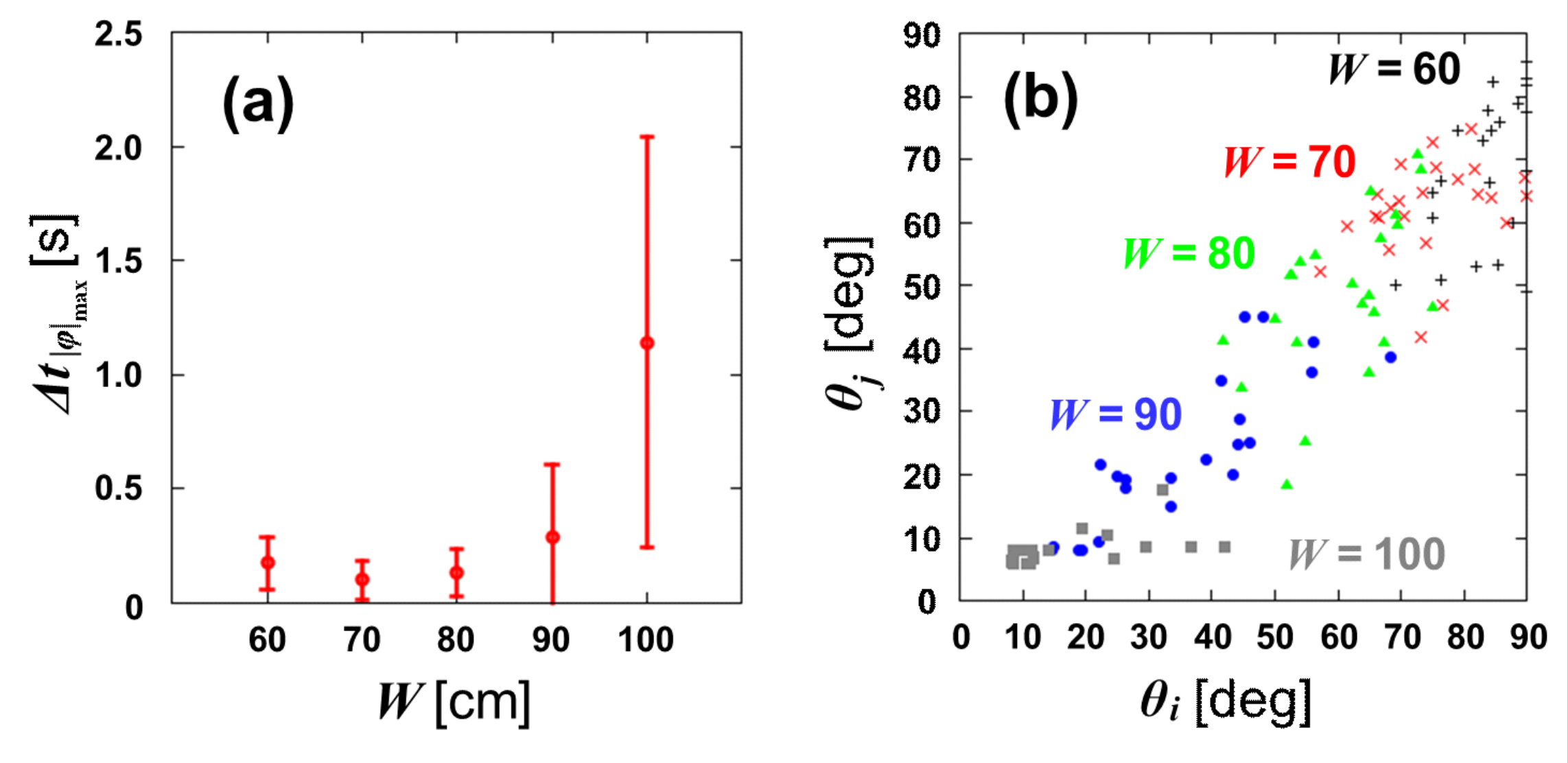}{16}{
(a) Average time gap $\Delta t _{|\varphi|_{\rm max}}$ between the achievement of the maximum rotational angles $|\varphi|_{{\rm max}, i}$ and $|\varphi|_{{\rm max}, j}$ $(i,j = 1, 2, 3, 4,\ i \not = j)$ before conversion with sample standard deviation.
(b) Pairs of passing rotational angles $(\theta_i, \theta_j)$ obtained from the experimental data after conversion
for various values of $W$.
The black crosses ($+$), red crosses ($\times$), green triangles ($\triangle$), blue circles ($\circ$), and gray squares ($\square$) represent the experimental data for $W=60$, 70, 80, 90, and 100 cm, respectively.
Note that the experimental data were plotted such that $\theta_i \geq \theta_j$ is satisfied.
}}

Figure \ref{figure6}a shows the averages and sample standard deviations of $\Delta t _{|\varphi|_{\rm max}}$ as functions of $W$ before conversion.
From this figure, it can be seen that the values are small for $W=60$, 70, and 80 cm, increasing slightly for $W=90$ cm, and increasing greatly for $W=100$ cm. 
Thus, a few conversions are necessary for $W=90$ cm and many conversions are necessary for $W=100$ cm.

Actually, a large $\Delta t _{|\varphi|_{\rm max}}$ was obtained when no clear maximum rotational angle was observed, as shown in Figs. \ref{fig:figure5}e-\ref{fig:figure5}h.
Therefore, $\Delta t _{|\varphi|_{\rm max}}$ can be used for judging not only whether the assumption $\theta_i = |\varphi|_{{\rm max}, i}$ is valid but also whether collision avoidance occurs with body rotation.
Judgment based on the values of the time gap is much easier than that based on the plots of rotational angle evolution, as shown in Fig. \ref{fig:figure5}.
Thus, the time gap $\Delta t _{|\varphi|_{\rm max}}$ was considered to be a simple and useful indicator for detecting collision avoidance with body rotation.

Figure \ref{figure6}b shows the pairs of passing rotational angles $(\theta_i, \theta_j)$ $(i \not = j)$ in the experimental data after conversion.
It can be seen that the passing rotational angles increased with decreasing corridor width.
This result is intuitively agreeable.
Moreover, the passing rotational angles were not always equal within a given pair, i.e., $\theta_i \not = \theta_j$ in general. 
Therefore, the rotation of a pair, as opposed to that of an individual, was considered.

\begin{figure}[t]
\centering
\includegraphics[width=8cm]{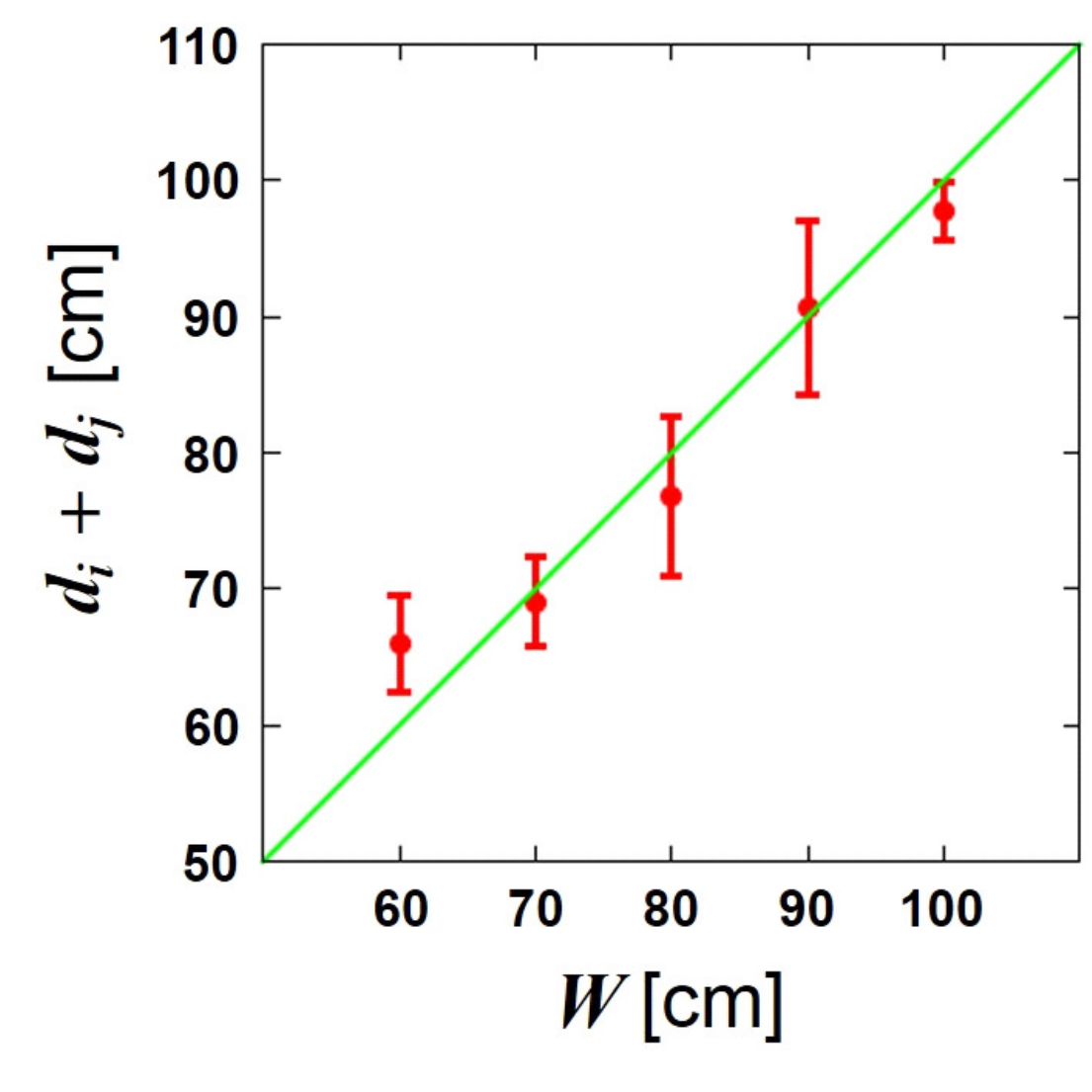}
\caption{
Sum of widths of the two ellipses $d_i+d_j$ as a function of corridor width $W$.
The green diagonal line represents the theoretical assumption (\ref{eq:W1}), i.e., $d_i+d_j=W$.
The red markers are the average values calculated using the least squares method by exploiting the experimental data and (\ref{eq:d}).
The error bars show the sample standard deviations.
}
\label{fig:fig_result_theoexp}
\end{figure}


\SUB{Validation of elliptic pedestrians}{Comparison}
To validate elliptic pedestrians and assumption 
(\ref{eq:W1}) during passing, 
the parameters of elliptic pedestrians were determined using the experimental data.

As discussed in Sec. \ref{PRAngles} and shown in Fig. \ref{figure6}b, the passing rotational angles of the two participants were generally different.
Therefore, it was necessary to investigate the relationship between $(\theta_i, \theta_j)$ and $W$.
Hence, the relationship between $d_i + d_j$ (the sum of the widths of the two ellipses) and $W$ was studied; this is much more appropriate in terms of comparison, because $d_i + d_j$ is one-dimensional, and its physical dimension is equal to that of $W$.

Since the participants were all male adolescents with similar builds, the semi-major and semi-minor axes of the elliptic pedestrians were assumed to be identical, i.e., $a_i =a$ and $b_i =b$ for $i=1$, 2, 3, and 4.
Then, (\ref{eq:d}) and (\ref{eq:W1}) were used along with the experimental passing rotational angles and the semi-major and semi-minor axes of the elliptic pedestrians ($a$ and $b$) were determined to minimize the following function (the least squares method):
\EQL{eq:leastSquareMethod}{
f(a,b) = \sum_{\rm{all\ data}} \LL W -  d_i(\theta_i, a, b) - d_j(\theta_j, a, b) \LR^2.
}
As a result, the parameters $(a,b)=$(24.9 cm, 15.5 cm) were obtained.
Then, the mean and sample standard deviation of $d_i+d_j$ for each $W$ were computed by substituting the calibrated semi-major and semi-minor axes $(a, b)$ and the experimental passing rotational angles $(\theta_i, \theta_j)$ into (\ref{eq:d}).
Figure \ref{fig:fig_result_theoexp} shows the mean and sample standard deviation of $d_i+d_j$ with the red markers and error bars, respectively. 

In Sec. \ref{MODEL}, it was assumed that pedestrians passing each other satisfied (\ref{eq:W1}), the green diagonal line in Fig. \ref{fig:fig_result_theoexp}.
Thus, the correspondence between the red markers and green diagonal line indicates the validity of this assumption.
It can be observed that the markers agree favorably with the diagonal line, namely, the error bars intersect with the green diagonal line except when $W=60$ cm. 

The discrepancy at $W=60$ cm is considered to occur as follows.
The walls of the experimental courses were constructed out of cardboard boxes, whose heights (62 cm) were much shorter than those of the participants (Fig. \ref{fig:figure3a}), so that the participants' upper bodies were allowed to pass over the corridor.
Therefore, it was considered that the participants pushed their shoulders out of the corridor and passed each other when the corridor width was narrow ($W=60$ cm) and collision avoidance was not easy; 
this decreased the passing rotational angles and increased $d_i+d_j$, as can be seen at $W=60$ cm in the plot of Fig. \ref{fig:fig_result_theoexp}.

The validity of the parameters $(a,b)=$(24.9 cm, 15.5 cm) determined from the experimental data was also examined.
Since $(a,b)$ are the semi-major and semi-minor axes of the ellipse, respectively, the effective shoulder width and bust depth of a model pedestrian correspond to  $2a=49.8$ cm and $2b=31.0$ cm, respectively.
In the passing scenario of the experiment, the shoulders of the participants did not come in contact with one other.
This is ascribed to the fact that pedestrians tend to maintain some distance from other pedestrians to avoid physical contact.
Therefore, it can be assumed that the physical size of the participants is smaller than the effective size obtained from the experimental data.
It has been reported that the average shoulder width and bust depth of Japanese men of age 20-24 years (without clothes) are 44.9 cm and 20.1 cm, respectively \cite{sizeJPN2004-2006}.
Considering clothing thickness and the fact the participants did not come into contact with each other, the statistical value of shoulder width can be considered to be appropriately smaller than the calibrated value.
However, the statistical value of the bust depth was approximately 10 cm smaller than the calibrated value.
This is likely because pedestrians are more hesitant about chest-to-chest contact than shoulder-to-shoulder contact.
Hence, the calibrated size of the elliptic pedestrians was concluded to be appropriate representations of the effective size of real pedestrians.

\SEC{Analysis of deceleration during passing in corridor experiment 1}{Deceleration}

This section elaborates upon the deceleration effect during passing.
Figure \ref{figure8} shows the average time for traveling 2 m ($T_{\rm{travel}}$) as a function of $W$ in course 1 of corridor experiment 1.
It can be seen that the qualitative characteristic of the plot changed greatly at $W=100$ cm, similar to the case of body rotation.
Specifically, the travel time was small and constant for $W=100$-140 cm, for which few clear body rotations were observed. 
By contrast, the travel time increased as $W$ decreased from 100 to 60 cm, for which the participants clearly rotated their body to pass each other.
This increase in the travel time is due to deceleration during passing with body rotation.

\FIGT{\EPSW{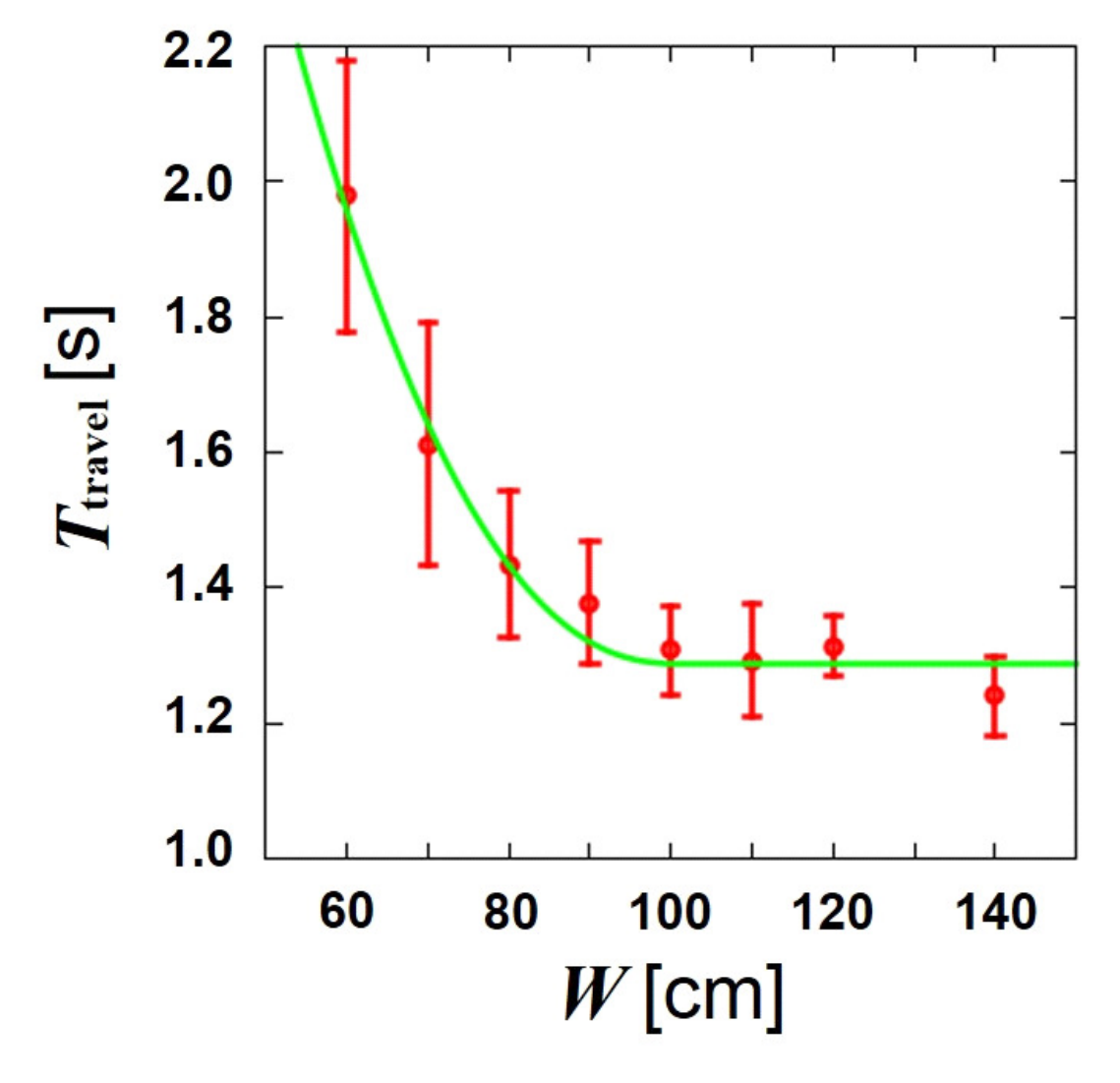}{8}{
Average travel time for 2 m ($T_{\rm{travel}}$) as a function of corridor width ($W$) in course 1 of corridor experiment 1.
The red markers and error bars represent the averages and sample standard deviations of the experimental results, respectively.
The green curve 
is depicted with (\ref{eq:Ttravel}) and (\ref{DeltaT}), whose parameters were calibrated with the experimental data.
}}

The travel time was considered to be represented by the sum of the travel time without deceleration $T_0$ and the increment due to deceleration during passing $\Delta T$, as follows:
\EQL{eq:Ttravel}{
T_{\rm travel} = T_0 + \Delta T(W),
}
where the explicit formulation of $\Delta T(W)$ was assumed as follows:
\EQL{DeltaT}{
\Delta T(W) = \LPM{ll}{
c_1 (W_{\rm cr} - W)^{c_2} & (W \le W_{\rm cr}),\\
0 & (W \ge W_{\rm cr}).
}}
$T_0 = 1.29$ s was calculated from the average travel time for $W=100$-140 cm, and $W_{\rm cr}$ was set to $W_{\rm cr} = 100$ cm.
Then, the parameters $c_1$ and $c_2$ were determined with the least squares method by using the average travel time for $W=60$-100 cm.
As a result, $c_1 = 1.94 \times 10^{-4}$ s/cm$^{c_2}$ and $c_2 = 2.21$ were obtained.
The green curve in Fig. \ref{figure8} is depicted with (\ref{eq:Ttravel}) and (\ref{DeltaT})
and agrees well with the experimental data.

$\Delta T(W)$ represents the increment of travel time during one pass with body rotation.
Thus, the influence of multiple ($N$ times) passes can be estimated as $\Delta T(W) \cdot N$, although more data are required to improve the quality of the result for real-world applications. 

\SEC{
Simulation model and comparison between experimental and simulation results
}{SIM}
\SUB{Simulation model for pedestrians with body rotation in a narrow corridor}{SimModel}

As described in Sec. \ref{INTRO}, pedestrians were represented by circles in many conventional models.
There are some models that represent pedestrians with ellipses and other more-realistic shapes; however, the body orientation and walking direction of a pedestrian are considered to be the same in these models.
Therefore, these previous models cannot reproduce collision avoidance by body rotation.
Thus, a new model was developed to simulate pedestrians' collision-avoidance behavior by body
rotation.

Figure \ref{fig_sim_1} shows a schematic of the model, the setup of which is equivalent to that of  corridor experiment 1 (explained in Sec. \ref{EXP}).
The length and the width of the corridor are $L$ and $W$, respectively.
There is a pedestrian R (L) at the left (right) end of the corridor who moves to the right (left) end in the simulation.

\FIGT{
\EPSW{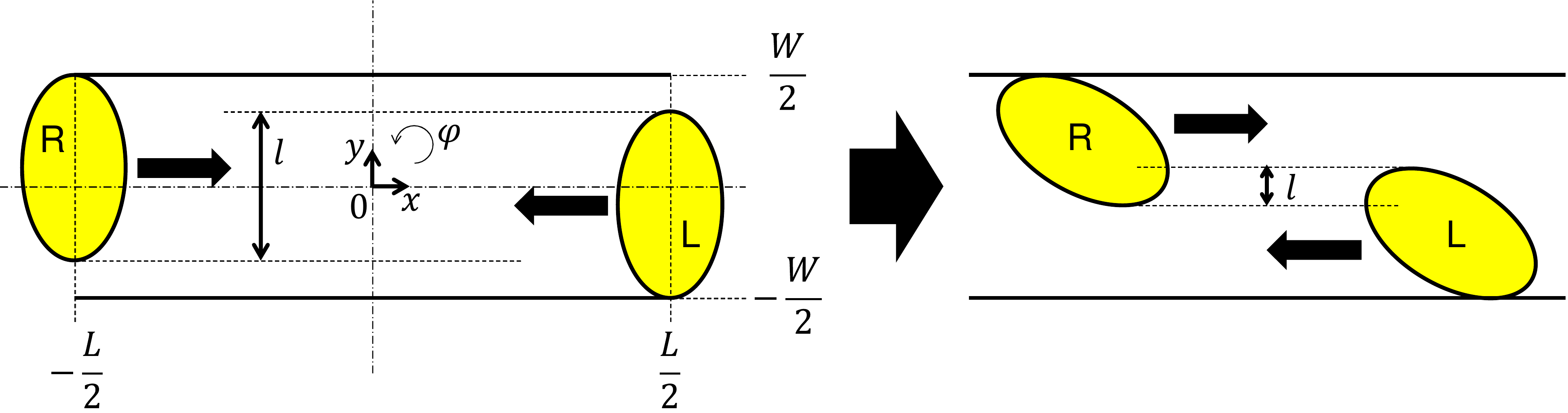}{16}{
Schematic of the simulation model.
Pedestrians R and L try to decrease the overlap length $l$ by evading and rotating according to (\ref{eqm_y1}) and (\ref{eqm_p1}), respectively.
}}

Pedestrians are represented with the same ellipses whose semi-major and semi-minor axes are $a$ and $b$, respectively, and their desired velocity is $v^{\rm des}_i$ $(i \in \{\rm{R}, \rm{L}\})$.
The position of the center and the rotational angle of pedestrians are described by $(x_i, y_i)$ and $\varphi_i$, respectively.

When the distance between the two pedestrians is larger than the threshold distance, i.e., $x_{\rm L} - x_{\rm R} > d_{\rm th}$, the two pedestrians do not interact with each other and keep walking with their desired velocities $v^{\rm des}_i$.
When the two pedestrians approach each other, i.e., $x_{\rm L} - x_{\rm R} \le d_{\rm th}$, they start to interact.

The pedestrians were considered to evade in the perpendicular direction $(y)$ to their moving direction $(x)$ and rotate their bodies to decrease the overlap length $l$ and avoid each other, as shown in Fig. \ref{fig_sim_1}.
Thus, pedestrians gradually control their positions $y_i$ and rotational angles $\varphi_i$ proportional to the overlap length $l$, which is calculated by exploiting (\ref{eq:d}).
Furthermore, pedestrians' walking speeds become smaller during side-stepping.
Therefore, the equations of motion of pedestrians before passing are described as follows:
\EQL{eqm_x1}{
\frac{dx_i}{dt} = v^{\rm des}_{i} \cos(\varphi_i),
}

\EQL{eqm_y1}{
\frac{dy_i}{dt} = k^{\rm A}_{y} l \cdot {\rm sign}(y_i-y_j),
}

\EQL{eqm_p1}{
\frac{d\varphi_i}{dt} = k^{\rm A}_{\varphi} l,
}
where $k^{\rm A}_{y}$ and $k^{\rm A}_{\varphi}$ are sensitivity parameters for the overlap length, ${\rm sign}(z)$ gives the sign of the argument $z$, and $y_j$ $(j \in \{\rm{R}, \rm{L}\})$ is the position of the opposing pedestrian.

After passing, 
i.e., $x_{\rm L} - x_{\rm R} < -2b$,
pedestrians try to restore their attitudes and move as fast as possible.
It was assumed that the restoring behavior was proportional to the evading distance and rotational angle, 
so that the equations of motion of pedestrians after passing can be described as follows:
\EQL{eqm_x2}{
\frac{dx_i}{dt} = v^{\rm des}_{i} \cos(\varphi_i),
}

\EQL{eqm_y2}{
\frac{dy_i}{dt} = -k^{\rm R}_{y} (y_i - y^0_i),
}

\EQL{eqm_p2}{
\frac{d\varphi_i}{dt} = -k^{\rm R}_{\varphi} (\varphi_i - \varphi^0_i),
}
where $k^{\rm R}_{y}$ and $k^{\rm R}_{\varphi}$ are sensitivity parameters for the deviation from the initial position ($y^0_i$) and initial rotational angle ($\varphi^0_i$), respectively.

The origin of the coordinate system was set to the center of the corridor and the $x$ and $y$ axes and positive direction of $\varphi$ were defined as in Fig. \ref{fig_sim_1}.
At the beginning of the simulation, pedestrians stood to touch the wall, so that the initial positions and rotational angles of the pedestrians were $(x^0_{\rm R}, y^0_{\rm R}, \varphi^0_{\rm R}) = (-L/2, W/2-a, 0)$ and $(x^0_{\rm L}, y^0_{\rm L}, \varphi^0_{\rm L}) = (L/2, -W/2+a, 0)$, respectively.
Since the pedestrians R and L move to the right and left, respectively, $v^{\rm des}_{\rm R} > 0$ and $v^{\rm des}_{\rm L} < 0$.


\SUB{Comparison of passing rotational angle and travel time between experiment and simulation}{SimComp}

The semi-major and semi-minor axes of the pedestrians were set as $(a, b) = (24.9 {\rm \ cm}, 15.5 {\rm \ cm})$, as obtained from the analysis shown in Sec. \ref{COMP}.
Then, the desired velocity and the threshold distance were determined as $v^{\rm des}_{i} = 155 {\rm \ cm /s}$, $d_{\rm th} = 150 {\rm \ cm}$ from the travel times in the case $W=$ 100, 110, 120, and 140 cm and video data, respectively.
To calibrate the parameters of evasion and rotation ($k^{\rm A}_{y}$, $k^{\rm A}_{\varphi}$, $k^{\rm R}_{y}$, and $k^{\rm R}_{\varphi}$), it was necessary to consider three requirements.
First, the passing rotational angles of the experiment discussed in Sec. \ref{COMP} and those of the simulation should coincide with each other.
Second, the travel times of the experiment discussed in Sec. \ref{Deceleration} and those of the simulation should also coincide with each other.
Finally, the overlap length during passing should be minimized.
These three requirements are inconsistent.
For example, if the error in travel time is decreased, the overlap length during passing may increase.
In this paper, we first tried to minimize the error of the passing rotational angle and travel time between the experiment and simulation, and then confirmed that the overlap length during passing was small enough.

An error function related to the passing rotational angle and travel time was considered:
\EQ{
\sum_{W \in W_{\rm exp}} \LL ( \theta'_{\rm sim}(W) - \theta'_{\rm exp}(W) )^2 + ( T'_{\rm travel, sim}(W) - T'_{\rm travel, exp}(W) )^2 \LR,
}
where $W_{\rm exp} = \{$60, 70, 80, 90, 100, 110, 120, 140$\}$ cm and $X'_{\rm Y}$ ($X=\theta$, $T_{\rm travel}$, ${\rm Y}=$ exp, sim) is the scaled passing rotational angle and scaled travel time, respectively, of the experiment and simulation.
$\theta_{\rm exp}$ is the average of the plots in Fig. \ref{figure6}b and $T_{\rm travel, exp}$ is the data plotted in Fig. \ref{figure8}.
Scaling was performed by the following formulation:
\EQ{
X'_{\rm Y} = \frac{ X_{\rm Y} - X_{\rm exp, min} } { X_{\rm exp, max} - X_{\rm exp, min} }.
}
Then, the parameter set that minimizes the error function from $k^{\rm A}_{y}$, $k^{\rm A}_{\varphi}$, $k^{\rm R}_{y}$, $k^{\rm R}_{\varphi}$ $\in [0, 10]$ in 0.1 increments was searched, and the parameters were set as in the caption of Fig. \ref{fig_sim_2}.
Under this condition, the maximum overlap length during passing was 4.2 cm, which is much smaller than the size of the pedestrians, and thus the parameter set was used. 

\FIGT{
\EPSW{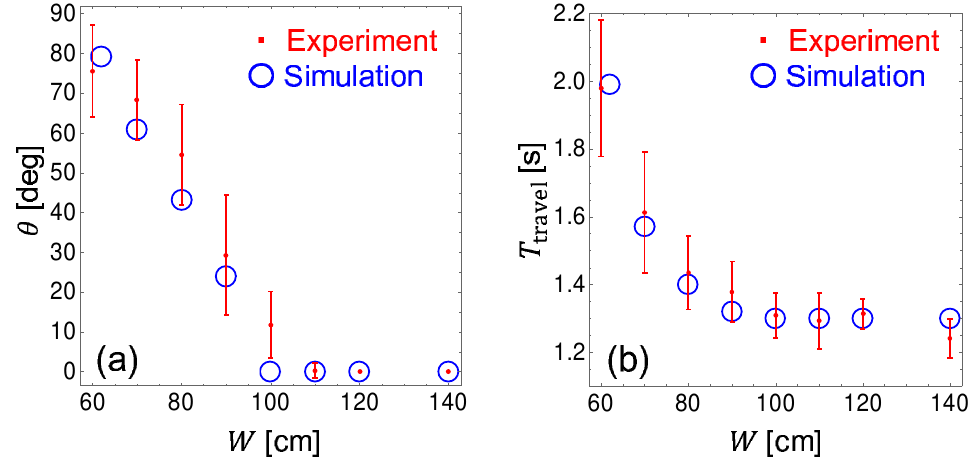}{16}{
(a) Passing rotational angles $\theta$ (b) travel time for 2 m ($T_{\rm{travel}}$) in the experiment and the simulation as functions of the corridor width $W$.
The red markers and error bars represent the averages and sample standard deviations of the experimental results, respectively.
The blue circles represent the simulation results.
The parameters in the simulation are $(a, b) = (24.9 {\rm \ cm}, 15.5 {\rm \ cm})$,
$(v^{\rm des}_{\rm R}, v^{\rm des}_{\rm L}) = (155 {\rm \ cm/s}, -155 {\rm \ cm/s})$,
and $(k^{\rm A}_{y}, k^{\rm A}_{\varphi}, k^{\rm R}_{y}, k^{\rm R}_{\varphi}) = (9.0 {\rm \ cm/(cm\cdot s)}, 6.0 {\rm \ deg/(deg\cdot s)}, 5.0 {\rm \ cm/(cm\cdot s)}, 7.0 {\rm \ deg/(deg\cdot s)})$.
Equations (\ref{eqm_x1})-(\ref{eqm_p2}) were numerically calculated by the Euler method with $\Delta t = 0.01$ s in the simulation.
}}

Figure \ref{fig_sim_2}a shows the passing rotational angle $\theta$ in the experiment and simulation as functions of the corridor width $W$.
The data plotted in Fig. \ref{figure6}(b) were averaged and used for the experimental results of $W=60$, 70, 80, 90, and 100 cm.
Similar data were also used for $W=110$, 120, and 140 cm.
The maximum $\varphi$ were plotted for the simulation result.

It can be observed that the simulation successfully reproduced the experimental result; however, the error is relatively large at $W=100$ cm.
This is due to the oscillation during walking in the experiment.
The pedestrians in the simulation do not oscillate during walking, so that $\varphi = 0$ throughout the simulation if $W \ge 2 \times 2a$ (=99.6) cm.

Note that the minimum corridor width in the simulation is not $W=60$ cm but $W=62$ cm, since the sum of the minimum widths of two pedestrians is $2b \times 2 = 62.0$ cm in the simulation.
As discussed in Sec. \ref{Comparison}, the pedestrians could push their shoulders out of the corridor due to the short wall in the experiment.
However, such behavior is not considered in the simulation, 
and thus it is very difficult for the pedestrians to pass each other in the corridor of $W=60$ cm in the simulation.

Figure \ref{fig_sim_2}b shows the travel times $T_{\rm travel}$ in the experiment (same as in Fig. \ref{figure8}) and simulation as functions of the corridor width $W$.
It can be seen that the two results are in agreement.

The above discussion confirms that the proposed model can simulate pedestrians passing in a narrow corridor in terms of passing rotational angle, travel time, and overlap length during passing.

\SUB{Bidirectional-flow simulation and fundamental diagram}{SimFD}

\FIGT{
\EPSW{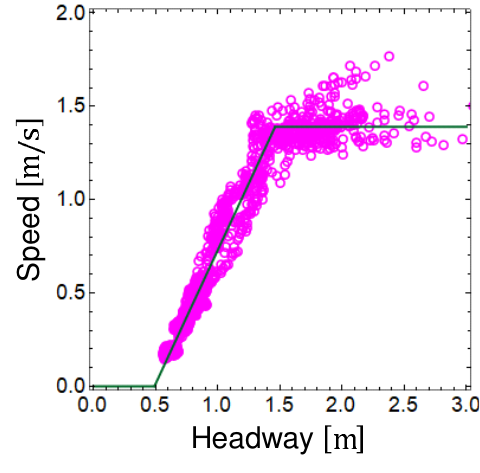}{8}{
Speed-headway relation in unidirectional flow.
The magenta circles and green plot represent the result of the experiments in \cite{Holl2016phd, Cao2017jstat, Juelich2013uni} and the calibrated desired-speed function $|v^{\rm des}| = s(h)$ (\ref{eq:piecewise}), respectively.
The parameters in $s(h)$ are set as $s^{\rm max}$ = 139 cm/s, $h_0$ = 49 cm, and $h_1$ = 146 cm.
}}

In this subsection, we conducted bidirectional-flow simulations with more than two pedestrians in a narrow corridor and depicted fundamental diagrams to validate the proposed model for crowd simulation.
To achieve this, the speed-headway relation should be modeled in addition to our body-rotation model.
We assumed that the desired-speed $|v^{\rm des}|$
\footnote{
The word ``desired-speed'' is usually used to describe the maximum speed without any obstruction; however, in this context, we used it as the speed determined by the headway distance without body rotation to maintain the consistency of the meaning of the word ``desired'' from the previous sections.
}
 is given by the following piecewise-linear function:
\EQL{eq:piecewise}{
|v^{\rm des}| = s(h) = \LPM{ll}{
0 & (0 \le h \le h_0), \\
\displaystyle s^{\rm max} \frac{h - h_0}{h_1 - h_0} & (h_0 \le h \le h_1), \\
s^{\rm max} & (h_1 \le h),
}
}
where the argument $h$ is the headway distance.
The maximum speed $s^{\rm max}$ [cm/s], $h_1$ [cm], and $h_2$ [cm] are the parameters of the desired-speed function.
We exploited the experimental data of unidirectional flow in \cite{Holl2016phd, Cao2017jstat, Juelich2013uni}
and determined these three parameters through the least squares method.
The results were $s_{\rm max}$ = 139 cm/s, $h_0$ = 49 cm, and $h_1$ = 146 cm.
Figure \ref{fig_sim_3_fitting} shows the plots of the experimental data and the desired-speed function (\ref{eq:piecewise}), which were confirmed to agree well with each other ($R^2$ = 0.73).

By using the calibrated desired-speed function (\ref{eq:piecewise}), we performed the unidirectional-flow simulations in the periodic corridor (circuit), whose length and width were $L$ = 1000 cm and $W$ = 50 cm, respectively.
We controlled the number of pedestrians $N$ from 1 to 17 by 1 and positioned them at equal intervals in the corridor at the beginning of the simulations.
Overtaking was not considered, therefore, evading or body-rotational behaviors were not observed.
Thus, the desired-speed function (\ref{eq:piecewise}) dominated the system.
As we can see from Fig. \ref{fig_sim_4_fd}, the experimental and simulation results agree well each other in both the speed-density (Fig. \ref{fig_sim_4_fd}a) and flow-density (Fig. \ref{fig_sim_4_fd}b) relations as well as the speed-headway relation (Fig. \ref{fig_sim_3_fitting}).

\FIGT{
\EPSW{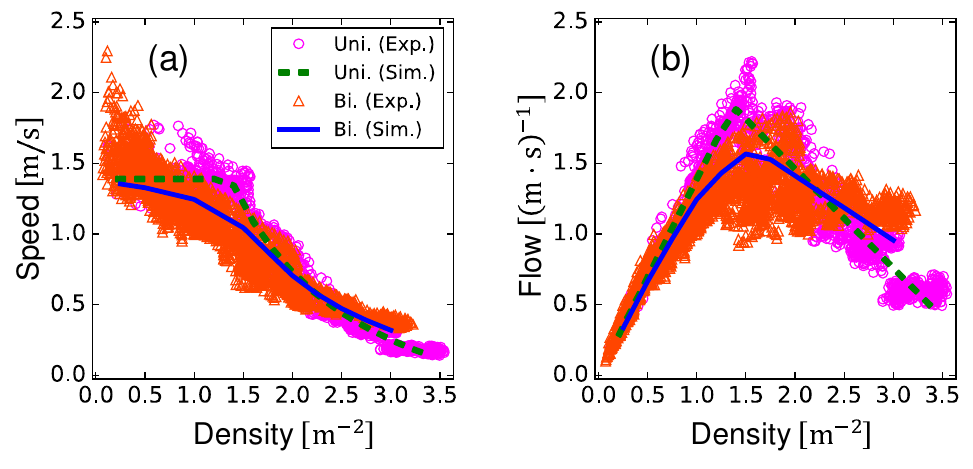}{16}{
Fundamental diagrams. (a) Speed-density relation. (b) Flow-density relation.
The magenta circles, green dashed curve, orange triangles, and blue solid curve represent the result of the unidirectional-flow experiment, unidirectional-flow simulation, bidirectional-flow experiment, and bidirectional-flow simulation, respectively.
The corridor length $L$ = 1000 cm, and the corridor width $W$ = 50 and 80 cm in the unidirectional and bidirectional-flow simulations, respectively.
The maximum speed $s^{\rm max}$ = 139 cm/s.
Other parameters are the same as in Figs. \ref{fig_sim_2} and \ref{fig_sim_3_fitting}.
Each simulation curve agrees well with the corresponding experimental plot.
}}

Then, we conducted bidirectional-flow simulations.
The width of the corridor was set as $W$ = 80 cm, and the number of total pedestrians $N$ was controlled from 2 to 24 by 2.
The number of the right- and left-going pedestrians were the same ($N/2$).
Both types of pedestrians were positioned at equal intervals in the corridor at the beginning of the simulations.
The sum of the shoulder widths of the two pedestrians $2 \times 2a$ = 99.6 cm was larger than $W$ = 80 cm, therefore, the pedestrians needed to evade others and rotate to pass each other.
The results of the bidirectional-flow simulations are presented with those of experiments in \cite{Holl2016phd, Cao2017jstat, Juelich2013bi} in Fig. \ref{fig_sim_4_fd}.
We can see that the simulation results are in agreement with the experimental results.

By comparing the experimental results of the unidirectional and bidirectional flows, we can observe that unidirectional flow achieves larger values of speed and flow than bidirectional flow when the density is smaller than the critical density $\approx$ 2.3 m$^{-2}$.
When the density is greater than the critical density, the values of bidirectional flow become larger than those of unidirectional flow.
These characteristic phenomena were successfully reproduced in our simulation results.

In the low-density situation, pedestrians can move freely in unidirectional flow; however, they have to interact with opponent pedestrians to avoid them in bidirectional flow.
Thus, speed and flow are greater in unidirectional flow than in bidirectional flow.

In the high-density situation, interactions with other pedestrians are unavoidable in both flows.
Overtaking is difficult in unidirectional flow because pedestrians cannot see behind themselves.
It is difficult for them to give way to fast followers.
Thus, the fundamental diagrams are mainly dominated by the simple speed-headway relation.
In contrast, pedestrians can see opponent pedestrians, give way, and pass by each other through evasion and rotation in bidirectional flow.
Due to these avoidance behaviors, speed and flow of bidirectional flow remain larger than those of unidirectional flow in the high-density situation.

Our simulation model succeeded in reproducing this phenomenon by introducing avoiding behaviors, i.e., evasion to the perpendicular direction and body-rotation.
Further, we would like to mention that the introduction of evasion alone was insufficient for bidirectional-flow simulations in a narrow corridor.
Since the sum of the shoulder widths of the two pedestrians $2 \times 2a$ = 99.6 cm was larger than the corridor width $W$ = 80 cm, body rotations were necessary to avoid deadlocks.

\SUB{Computational time}{SimCTime}

When simulation models are extended to improve their accuracy, more computational time is often required.
Therefore, weight between modeling accuracy and computational time is important to enable the practical use of simulation models.
In this subsection, we investigated the computational time of bidirectional-flow simulations in a narrow corridor.

We compared four cases.
The first case is the same as the proposed model introduced in the former subsections, which includes both evasion in the perpendicular direction and body rotation (Evasion and rotation).
In the second case, only evading behavior was considered, i.e., the body-rotation effect was removed (Only evasion).
In the third case, only body-rotation behavior was considered, i.e., the evading effect was removed (Only rotation).
In the fourth case, both evasion and body-rotation effects were removed, i.e., pedestrians did not avoid each other (No avoidance).

In Evasion and rotation and Only rotation cases, pedestrians have to be represented by ellipses.
On the contrary, in Only evasion and No avoidance cases, there are no effects attributed to elliptic pedestrians.
In the first three cases (except No avoidance case), there are interactions between pedestrians, i.e., pedestrians detect other pedestrians (detecting method in our simulation works).

Note that we conducted the simulations in this subsection only for evaluating the computational time.
Thus, the latter three cases (except Evasion and rotation) failed to complete accurate simulation, i.e., pedestrians passed through each other in spite of the large overlap lengths.

\FIGT{
\EPSW{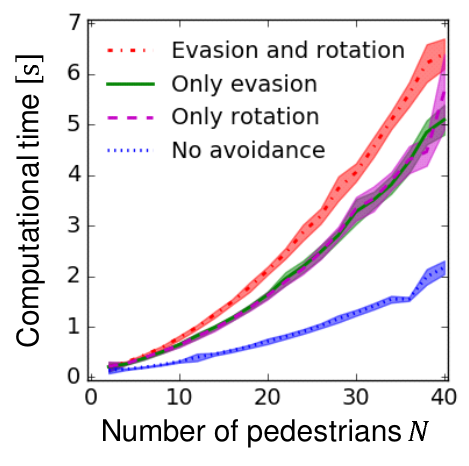}{8}{
Computational times of the simulations of bidirectional flow as functions of the total number of pedestrians $N$.
The red dashed-dotted, green solid, magenta dashed, and blue dotted curves represent the mean of Evasion and Rotation, Only evasion, Only, rotation, and No avoidance cases, respectively.
The filled area of each plot represents the sample standard deviation.
100 simulations were conducted for each $N$.
The parameters are the same as those in Fig. \ref{fig_sim_4_fd}.
The desktop computer with Windows 8.1 Enterprise, Intel(R) Core(TM) i7-4770 CPU @ 3.40GHz, and RAM 24.0 GB was used for the simulations.
}}

Figure \ref{fig_sim_5_ctime} shows the computational times of the four cases.
We see that the computational times increase as the number of pedestrians $N$ increases.
This type of power-law-like increase of computational time is common for force-based pedestrian models.
As we can imagine, the computational time of Evasion and rotation case is the largest, while that of No avoidance case is the smallest.
The computational times of Only evasion and Only rotation cases are similar and closer to that of Evasion and rotation case than No avoidance case.
This result indicates that the introduction of body-rotation behavior to the simulation models does increase the computational time; however, the increment becomes smaller if evading behavior has already been considered.
This is due to the cost of the detecting method.
When we introduce the first avoiding behavior (either evasion or rotation), we need to implement both the detecting and avoiding methods.
However, in the introduction of the second avoiding behavior, we do not need to add the detecting method anymore.
Therefore, the cost of introducing body-rotation behavior is not very high compared with the improvement of the accuracy if some interactions, which accompany the detecting method of other pedestrians, have already been considered in simulation models.

\SEC{
Setup and conditions of corridor experiment 2
}{EXP1-2}
In corridor experiment 1 described in Secs. \ref{EXP} and \ref{COMP}, all four participants were male adolescents.
Thus, the same experiment was performed once again with male and female participants of various ages to demonstrate that the obtained results are valid for all pedestrians.

The experiment was conducted in the lecture hall at RCAST, The University of Tokyo, Japan.
Four courses were constructed using cardboard boxes.
A single course was the same as that in corridor experiment 1 (Fig. \ref{fig:figure3a}).
There were five male participants, ages 21, 24, 29, 66, and 68, and four female participants, ages 36, 45, 61, and 64.
The average shoulder widths and bust depths of the male and female participants are summarized in Tab. \ref{tab:shoulderbust}.
The average shoulder widths of the males and females were significantly different in terms of statistics ($p<0.05$), while the bust depths were not.
The tablets for recording angular velocities were strapped on the participants.
We instructed the participants similarly as we did in corridor experiment 1.

The width $W$ of the two courses were varied as 60, 70, 80, 90, 100, 120, and 140 cm and the angular velocity of each participant was observed.
By changing the passing side (2 patterns) and participant pairing (combination choosing 2 participants from 9 = 36 patterns), 72 types of trials were executed for each $W$.
Thus, $72 \times 7$ (widths) = 504 datasets were obtained in total.

\begin{table}[t!]
\centering
\caption{
Average shoulder widths and bust depths of the male and female participants.
The values in the parentheses represent population standard deviations.
The average shoulder widths of the males and the females were significantly different ($p<0.05$), whereas the average bust depths were not.
Note that the average shoulder width and bust depth of the male participants are larger than the statistical values introduced in Sec. \ref{Comparison}.
This is likely due to clothing thickness.
}
\label{tab:shoulderbust}
\begin{tabular}{ccc}
\hline
Gender & Shoulder width [cm] & Bust depth [cm] \\ \hline
Male & 48.4 (3.7) & 26.0 (4.2) \\ \hline
Female & 41.0 (1.4) & 23.4 (2.4) \\ \hline
\end{tabular}
\end{table}

\SEC{
Analysis of passing rotational angles in corridor experiment 2
}{COMP1-2}


\SUB{Data processing}{Data_processing}

The time series of the angular velocities $\omega_i(t)$ ($i=0$, 1, 2, $\cdots$, and 8 (ID number)) were obtained at approximately 50 Hz from the tablets.
Then, the same data processing as in corridor experiment 1 was conducted.
The datasets of the angular velocities $\omega_i(t)$ were numerically integrated and the time series of the rotational angle of each participant $\varphi_{i}(t) \in [-180^{\circ}, 180^{\circ}]$ was obtained.
The time series of the pair that includes $\delta t_{i,n} \geq$ 0.1 s were discarded since it was impossible to compute the accurate rotational angles for such time series, as explained in Sec. \ref{TimeSeries}.
Actually, 4, 3, 5, 3, 2, 4, and 4 datasets were removed from $W=60$, 70, 80, 90, 100, 120, and 140 cm, respectively.
The passing rotational angles $\theta_i$ for $W=60$, 70, 80, 90, and 100 cm were calculated in a manner similar to that employed for corridor experiment 1, since no clear body rotations were observed for $W=120$ and 140 cm.
Any dataset that does not satisfy the conditions of the two thresholds ($\theta_i$ and $\Delta t _{|\varphi|_{\rm max}}$) is converted appropriately before being used for analysis, as described in \ref{100cm}.


Figure \ref{fig_1-2_all}a shows the pairs of passing rotational angles $(\theta_i, \theta_j)$ $(i \not = j)$ in the experimental data after conversion.
The passing rotational angles can be observed to increase with decreasing corridor width, as was also observed in corridor experiment 1.

\FIGT{\EPSW{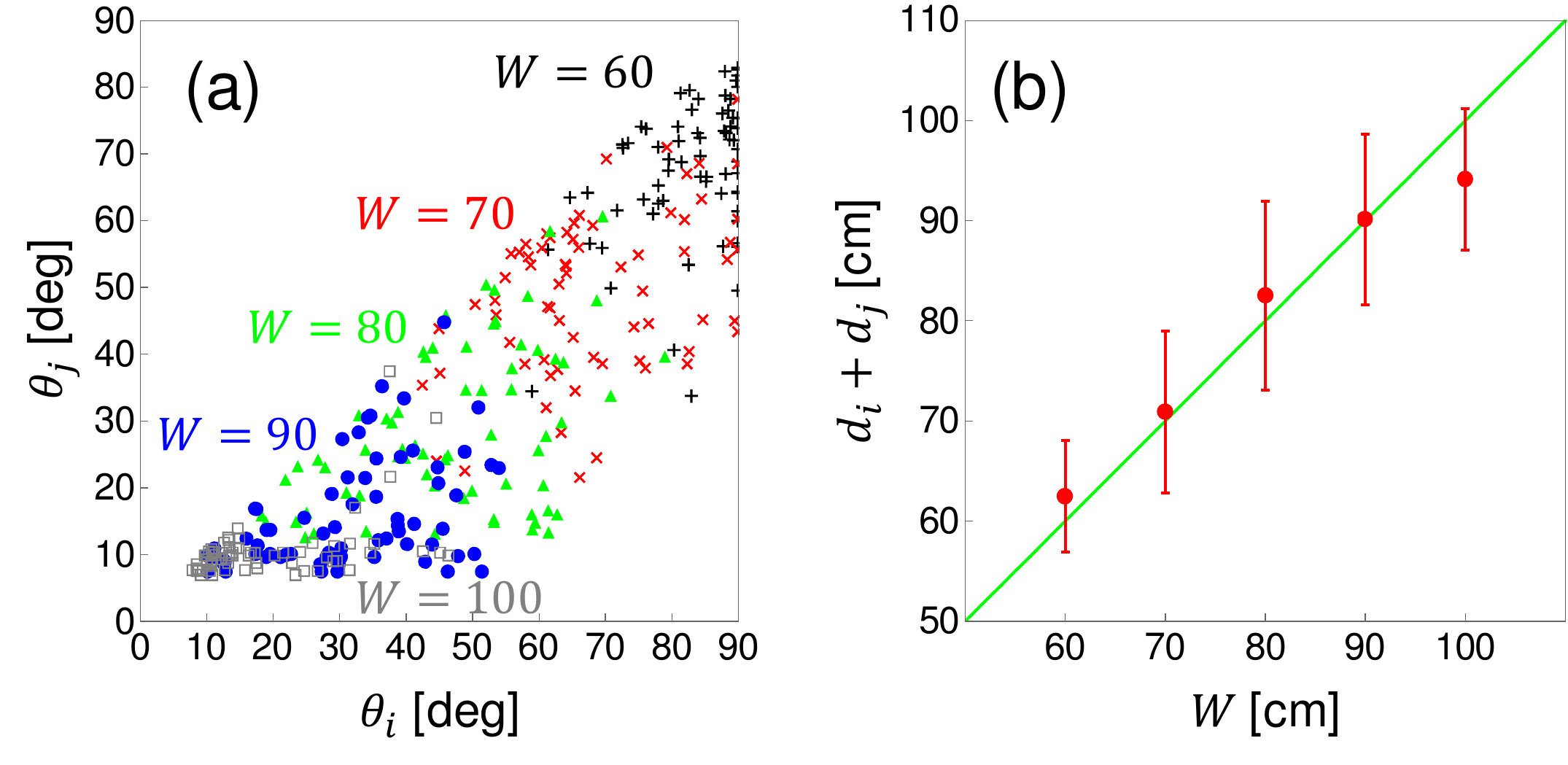}{16}{
(a) Pairs of passing rotational angles $(\theta_i, \theta_j)$ obtained from the experimental data after conversion for various values of $W$.
The black crosses ($+$), red crosses ($\times$), green triangles ($\triangle$), blue circles ($\circ$), and gray squares ($\square$) represent the experimental data for $W=60$, 70, 80, 90, and 100 cm, respectively.
Note that the experimental data are plotted such that $\theta_i \geq \theta_j$ is satisfied.
(b) Sum of the widths of the two ellipses $d_i+d_j$ as a function of corridor width $W$.
The green diagonal line represents theoretical assumption (\ref{eq:W1}), i.e., $d_i+d_j=W$.
The red markers are the average values calculated using the least squares method by exploiting the experimental data and (\ref{eq:d}).
The error bars show the sample standard deviations.
}}


\SUB{Validation of elliptic pedestrians}{Comparison1-2}
To validate elliptic pedestrians and assumption (\ref{eq:W1}) during passing, regardless of gender and age, a calculation similar to that for corridor experiment 1 was conducted.
Since the shoulder width and bust depth of each participant was measured, error parameters for the semi-major axis $\epsilon_a$ and semi-minor axis $\epsilon_b$ were introduced to minimize the following function (the least squares method):
\EQL{eq:leastSquareMethod1-2}{
g(\epsilon_a,\epsilon_b) = \sum_{\rm{all\ data}} \LL W -  d_i(\theta_i, a_i+\epsilon_a, b_i+\epsilon_b) - d_j(\theta_j, a_j+\epsilon_a, b_j+\epsilon_b) \LR^2,
}
where $a_i$ $(a_j)$ and $b_i$ $(b_j)$ are the half of the shoulder width (semi-major axis) and bust depth  (semi-minor axis) of participant $i$ $(j)$, respectively.
As a result, $(\epsilon_a, \epsilon_b)=$(1.6 cm, 2.1 cm) were obtained, which are positive small values.
Thus, it can be considered that $a_i +\epsilon_a$ ($a_j +\epsilon_a$) and $b_i +\epsilon_b$ ($b_j +\epsilon_b$) represent 
the effective shoulder width and bust depth, respectively, as the participants did not come in contact with each other throughout the experiment.
Hence, the elliptic representation of pedestrians and assumption (\ref{eq:W1}) are validated for both males and females of various ages.
Furthermore, as  $\epsilon_a < \epsilon_b$, it was reconfirmed that pedestrians hesitate to a greater extent to come into contact with other pedestrians at their chest than at their shoulders.

The mean and sample standard deviation of $d_i+d_j$ were computed for each $W$ by substituting the half of the shoulder width and bust depth $(a_i, b_i)$, $(a_j, b_j)$, 
calibrated parameters $(\epsilon_a, \epsilon_b)$,  and experimental passing rotational angles $(\theta_i, \theta_j)$ into (\ref{eq:d}).
Figure \ref{fig_1-2_all}b shows the mean and sample standard deviation of $d_i+d_j$ with the red markers and error bars, respectively. 
The correspondence between the red markers and green diagonal line also indicates the validity of this assumption.
Some discrepancy between the red marker and green diagonal line can be observed at $W=100$ cm.
As the female participants, whose average shoulder width was 41.0 cm as shown in Tab. \ref{tab:shoulderbust}, were included in the experiment, the value of $d_i+d_j$ decreases.


\FIGT{\EPSW{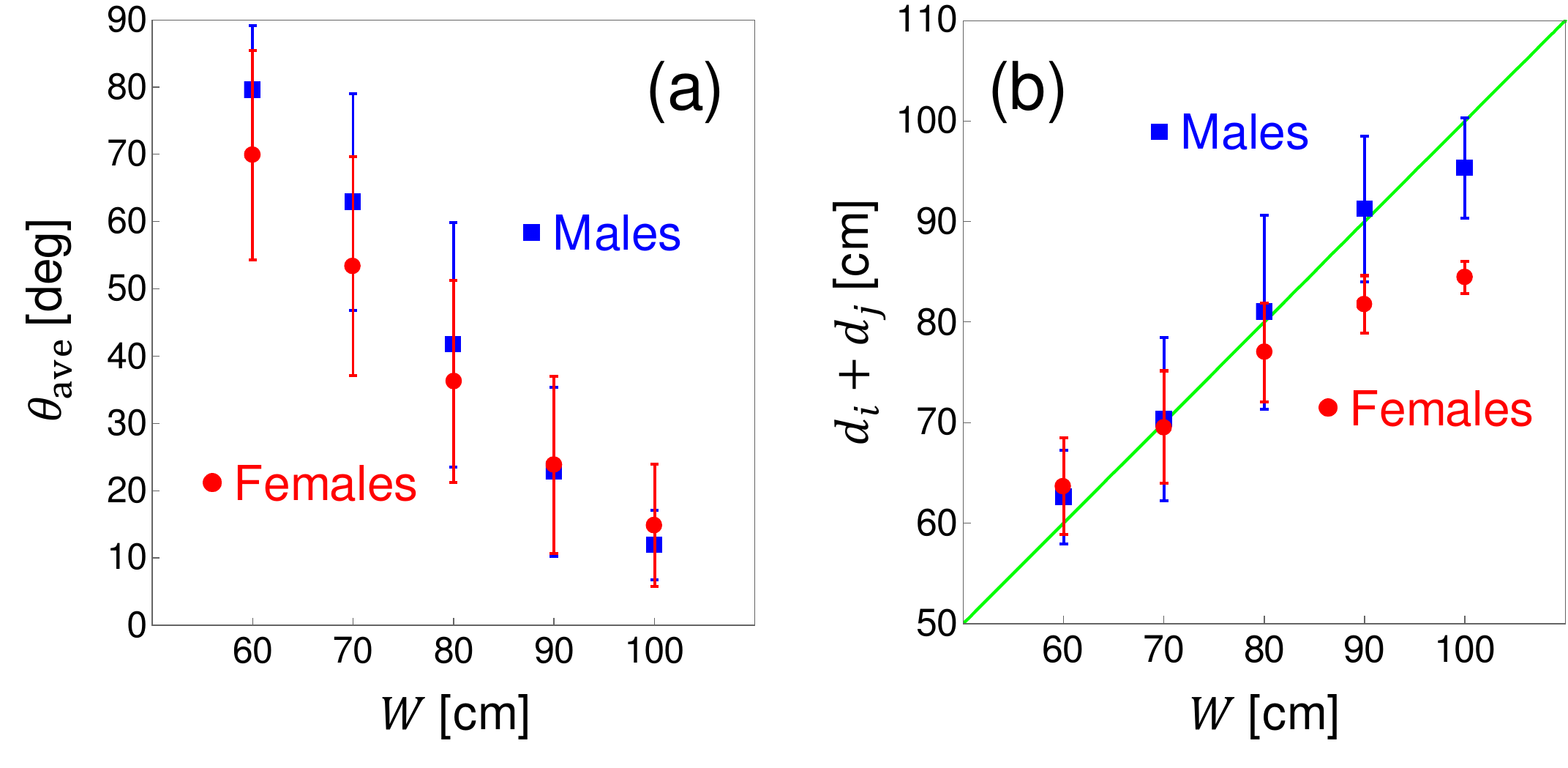}{16}{
(a) Average passing rotational angles as functions of the corridor width $W$ for the male and female pairs.
Blue squares and red circles represent the result of the male pairs and the female pairs, respectively.
The error bars show the sample standard deviations.
We see that the average rotational angles for the male pairs are significantly larger than those for the female pairs at $W=60$ and 70 cm ($p<0.05$).
(b) Sum of widths of the two ellipses $d_i+d_j$ as functions of the corridor width $W$ for the male and female pairs.
Blue squares and red circles represent the result of the male pairs and the female pairs, respectively.
The error bars show the sample standard deviations.
The green diagonal line represents theoretical assumption (\ref{eq:W1}), that is, $d_i+d_j=W$.
We see that the results of both the male and female pairs agree well with theoretical assumption (\ref{eq:W1}) when the corridor width $W$ is smaller than the double of the average shoulder widths described in Tab. \ref{tab:shoulderbust}.
}}

\SUB{Difference between the male and female pairs}{MaleFemale}
Since the average shoulder widths of the male and female participants were significantly different, the male and female pairs were analyzed separately in this section
\footnote{
There were no clear differences in the body sizes or passing rotational angles between the younger (adolescents and middle-aged adults) and older (elderlies) participants.
}.
Figure \ref{fig_1-2_men_women}a shows the average passing rotational angles $\theta_{\rm{ave}}$ as functions of the corridor width $W$.
The plots with the blue squares and red circles are the results of the male and female pairs, respectively.
The results of the male-female pairs, which were included in Fig. \ref{fig_1-2_all}, were removed from Figure \ref{fig_1-2_men_women}.
It can be seen that $\theta_{\rm{ave}}$ decreases as $W$ increases for both the male and female pairs.
Moreover, $\theta_{\rm{ave}}$ is significantly larger for the males than for the females at $W=60$ and 70 cm ($p<0.05$), respectively.
These differences can be attributed to the difference in the average shoulder width.
Since the shoulder widths of the males were significantly larger than those of the females, the males had to rotate more than the females.

The error parameters for the male $(\epsilon_{a, \rm{M}}, \epsilon_{b, \rm{M}})$ and female $(\epsilon_{a, \rm{F}}, \epsilon_{b, \rm{F}})$ pairs were also computed.
For the male pairs, the data for $W=60$-100 cm were used and $(\epsilon_{a, \rm{M}}, \epsilon_{b, \rm{M}})$ were calculated, which minimized (\ref{eq:leastSquareMethod1-2}) with the least square method.
For the female pairs, only the data for $W=60$-80 cm were used as the females did not need to rotate for $W=90$ and 100 cm; this is because the maximum sum of the shoulder width of the pair was smaller than 90 cm.
It was found that $(\epsilon_{a, \rm{M}}, \epsilon_{b, \rm{M}}) = (0.0\ \rm{cm}, 1.9\ \rm{cm})$ and $(\epsilon_{a, \rm{F}}, \epsilon_{b, \rm{F}}) = (1.2\ \rm{cm}, 3.0\ \rm{cm})$;
these values were small enough to be regarded as the difference between the effective and real body sizes.
$\epsilon_{a, \rm{M}} < \epsilon_{b, \rm{M}}$ and $\epsilon_{a, \rm{F}} < \epsilon_{b, \rm{F}}$, confirming once again that pedestrians hesitate 
to make chest contact with other pedestrians more so than shoulder contact.
Furthermore, since $\epsilon_{a, \rm{M}} < \epsilon_{a, \rm{F}}$ and $\epsilon_{b, \rm{M}} < \epsilon_{b, \rm{F}}$, it can be concluded that the females kept a greater distance between each other than the males in the passing situation in our experiment.

Figure \ref{fig_1-2_men_women}b shows the sum of the widths of the two ellipses $d_i + d_j$ as functions of the corridor width $W$ for the male and female pairs.
It can be observed that the results for the male pairs agree with the theoretical assumption (\ref{eq:W1}) for $W=60$-90 cm.
At $W=100$ cm, the result for the male pairs is smaller than the theoretical assumption (\ref{eq:W1}).
This small discrepancy can be explained as follows: since the average shoulder width of the males was 48.4 cm, as shown in Tab. \ref{tab:shoulderbust}, the males did not need to rotate for $W=100$ cm in most cases; however, pedestrians always rotated to some degree during walking (see \ref{100cm}).
Thus, the widths of the males reduced to below 100 cm, and the blue square was plotted below the green diagonal line at $W=100$ cm.
On the other hand, the results of the females agree with the theoretical assumption (\ref{eq:W1}) for $W=60$-80 cm.
As we mentioned in the second paragraph in this section, the maximum sum of the shoulder width of the female pairs was smaller than 90 cm, 
so that the discrepancies at $W=90$ and 100 cm do not harm our assumption.

\SEC{
Cross flow experiment
}{CROSSFLOW}
This section considers cross flow and investigates body-rotation behavior in a more complex and realistic scenario.
There are two motivations for studying cross flow in this paper.

The first motivation is to observe that pedestrians do step sideways, i.e., discrepancies between the body orientation and walking directions in cross flow.
Side-stepping was observed in the bidirectional-flow experiment in \cite{Jin2017}.
If such a phenomenon is observed in cross flow, which is more realistic than passage in a narrow corridor and more complex than bidirectional flow, it can be strongly confirmed once again that body rotation is an indispensable element in simulating pedestrian dynamics.

The second motivation is to understand the relationship between density and body rotation.
A narrow corridor necessitates body rotation; however, there are several options to avoid collisions in cross flow.
Although these additional options are beyond the scope of this paper, clarifying the relationship between density and body rotation would show when body-rotation behavior becomes important in simulations.

The experimental setup and conditions are discussed in Sec. \ref{CrossExp}, and the results that correspond to the two motivations presented above are discussed in Secs. \ref{CrossRes1} and \ref{CrossRes2}, respectively.


\SUB{Setup and conditions}{CrossExp}

\begin{figure}[t]
\centering
\includegraphics[width=10cm,clip]{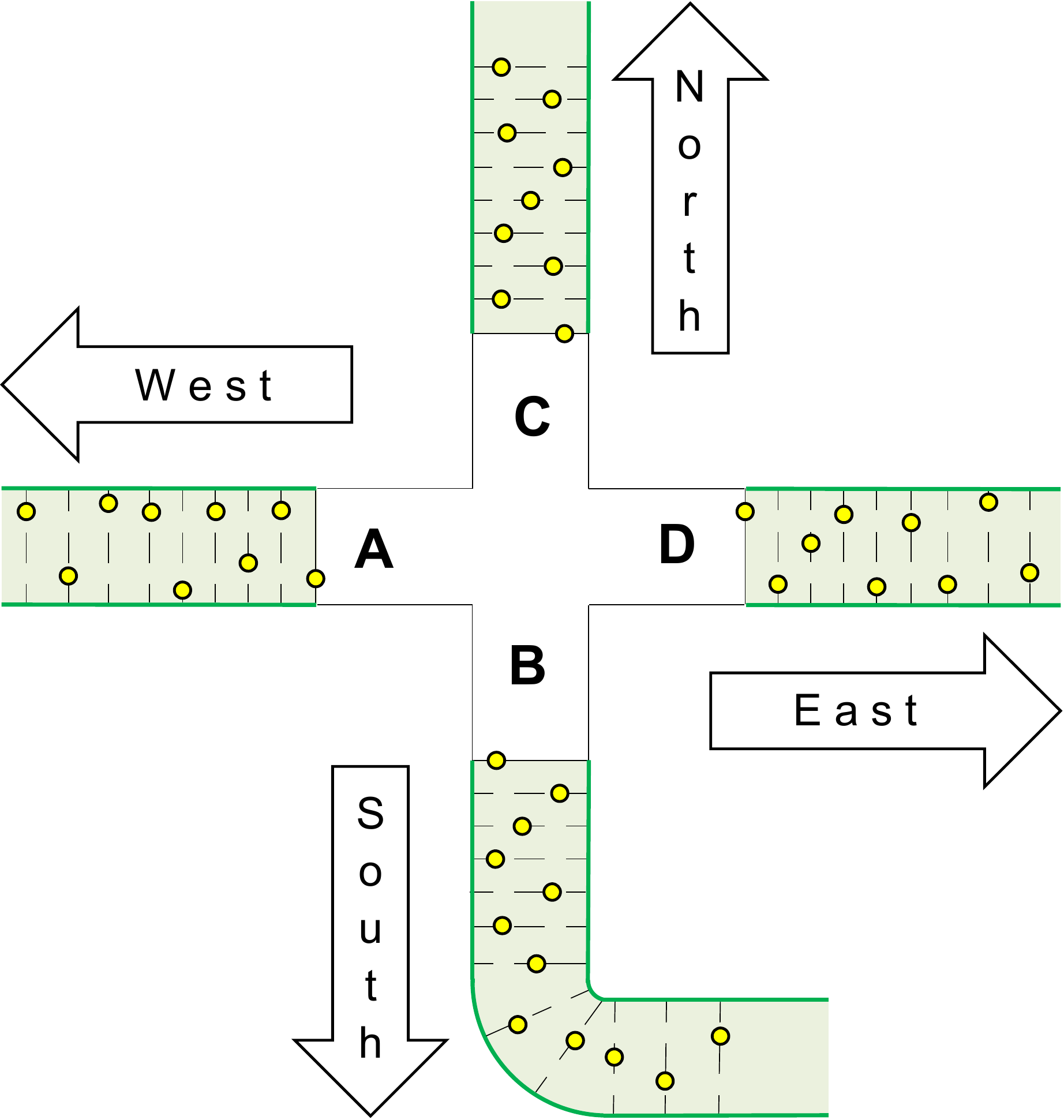}
\caption{
Geometrical configuration for the cross-flow experiment (at Execution 1). South leg had to be bent due to space limitations. Starting positions are given as dashed lines along the four legs. Participants were allowed to take any position along the line, but only one participant was allowed at each line. Configuration given here is only representative as participants took different positions during each repetition. 
}
\label{fig:cross_geometry}
\end{figure}

The experiment was conducted on the street in front of RCAST building 4 at The University of Tokyo, Japan.
An intersection, like the one shown in Fig.~\ref{fig:cross_geometry}, was constructed by creating four paths intersecting each other in a delimited area.
A video camera placed in an azimuthal position 20 m above the ground was used to record pedestrian motion in the intersecting section.
Due to space limitations, South leg had to be bent as shown in Fig.~\ref{fig:cross_geometry}.
Note that later data analysis (in particular concerning the flow in the four different directions \cite{Feliciani2017PhD}) revealed that the bending did not influence on participants' behaviors, making it possible to consider each leg independently.

Inside each leg, starting lines were drawn every 1 m longitudinally on the ground (the first starting line was 4 m from the center of the intersection).
The participants were allowed to take any position along the different starting lines, but only one participant was allowed per line.

In total, 50 male college students took part in the experiment and were divided into the four legs in groups of 12 or 13 participants (Group A, B, C, and D in Tab. \ref{tab:id_turning}).
As a consequence, unaligned lanes of respectively 11 and 12 m formed in the four legs.
An ID number was assigned to each participant and each one was asked to turn in a different direction at the intersection (details are given in Tab.~\ref{tab:id_turning}).
An almost equal portion of participants had to turn left and right.
Participants going straight were less than the ones turning.
The turning direction was given them using a small paper provided at the beginning of the experiment; therefore, no one was aware of the turning direction of others.

The participants started walking simultaneously toward the intersection when instructed to go.
Then, they turned according to the assigned directions at the intersection.
After leaving the intersection, the participants walked for a long distance to avoid the formation of congestion close to the exit.

\begin{table}[t!]
\centering
\caption{
Group and turning direction assigned to each participant.
Participants equipped with a tablet are given in bold.
}
\label{tab:id_turning}
\begin{tabular}{cccc}
\hline
Group & Turning right 							& Turning left 								& Going straight 				\\ \hline
A 		& \textbf{1},4,7,10						& \textbf{2},5,8,11						& \textbf{3},6,9,12 		\\ \hline
B 		& 13,15,18,21,\textbf{23}			& 14,16,19,22,\textbf{24}			& 17,20,25 							\\ \hline
C 		& 26,28,31,\textbf{35},37			& 27,29,32,\textbf{36}				& 30,33,34	 						\\ \hline
D 		& 39,41,44,\textbf{47},50			& 38,40,42,45,\textbf{48}			& 43,46,\textbf{49} 		\\ \hline
Total	& 19 (4 tablets)							& 18 (4 tablets)							& 13 (2 tablets)				\\ \hline
\end{tabular}
\end{table}

To avoid any possible learning process, in other words, to repeat the experiment under ``fresh'' conditions, the groups were switched after each repetition, as shown in Tab.~\ref{tab:start_switch}.
This switching strategy made it very unlikely that participants were able to memorize others' behaviors.
The participants' positions were also longitudinally switched at each repetition to avoid always having the same participants at the beginning or end of each leg.

\begin{table}[t!]
\centering
\caption{
Starting leg (West, South, North, East) for each group (A, B, C, D) during the different executions.
This strategy was chosen to avoid a possible learning resulting from the repeated executions.
Cardinal directions refer to Fig.~\ref{fig:cross_geometry}. (The same group locations are relative to Execution 1.)
}
\label{tab:start_switch}
\begin{tabular}{cccccc}
\hline
Execution	& West 		& South 	& North		& East	\\ \hline
1			& A			& B			& C			& D		\\ \hline
2			& A			& C			& B			& D		\\ \hline
3			& D			& C			& B			& A		\\ \hline
4			& D			& B			& C			& A		\\ \hline
\end{tabular}
\end{table}

The participants were given caps of different colors allowing us to precisely track their positions.
Moreover, as shown in Tab.~\ref{tab:id_turning}, some of the participants (10 in total) were equipped with a tablet to measure their bodies' rotational angles, as discussed in Sec. \ref{EXP}.
The participants equipped with tablets wore blue, red or green caps, whereas the rest of the participants wore yellow caps.
By knowing the combination of cap's color and starting leg, we were able to combine the body rotational angles and the positions (or trajectories) of each participant obtained from the tablets and the video camera, respectively.
This allowed the time at which a given participant crossed the intersection to be precisely determined and the data relative to that particular moment to be analyzed.

\begin{figure}[t]
\centering
\includegraphics[width=10cm,clip]{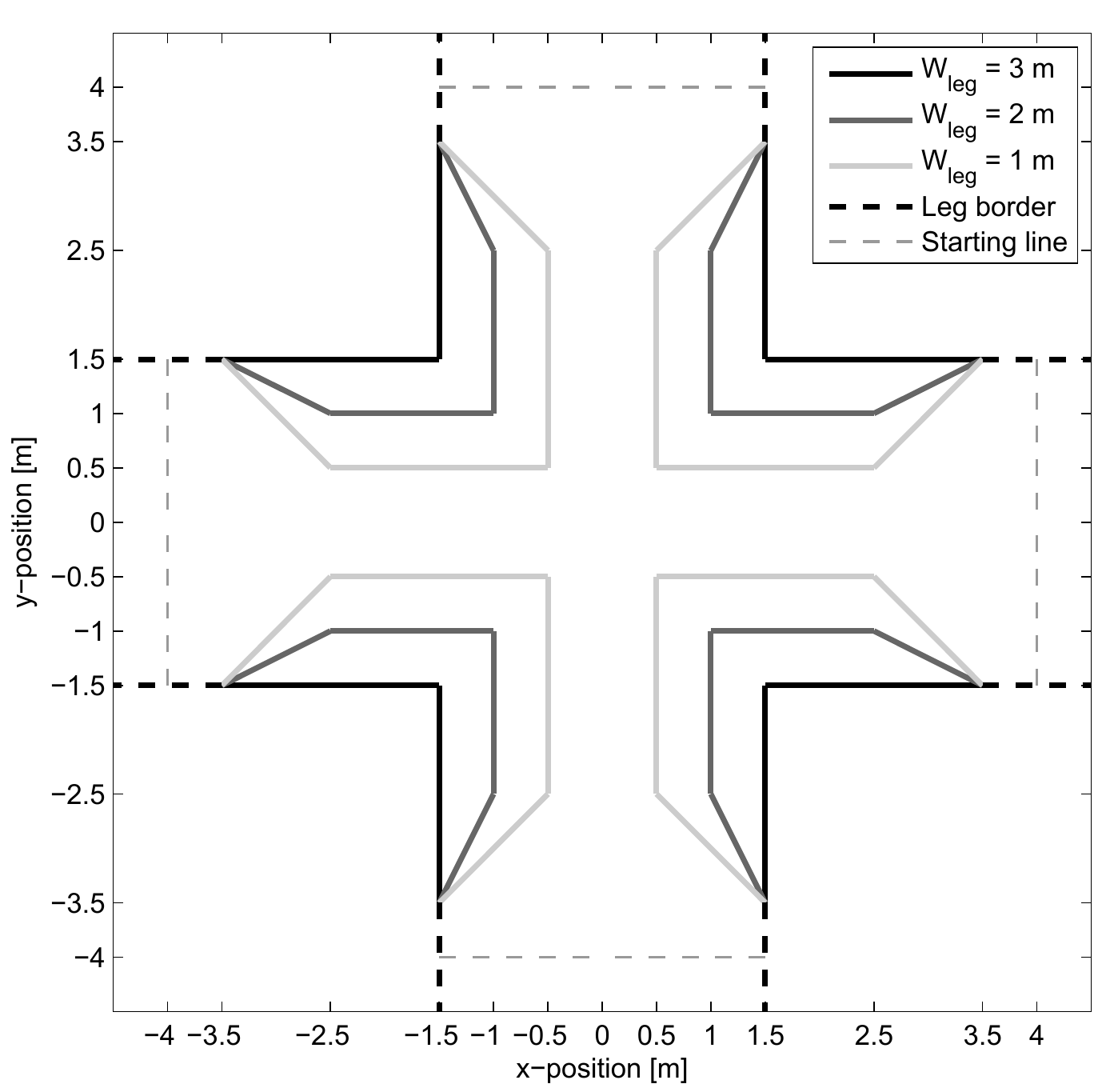}
\caption{
Configuration of the central part of the intersection for the leg width $W_{\rm{leg}}=3$, 2, and 1 m.
Width of the waiting areas in each legs were kept unchanged, while a bottleneck was created in each entrance to increase the density in the central part. 
}
\label{fig:section_width}
\end{figure}

\begin{figure}[t!]
\begin{tabular}{cc}
\begin{minipage}[t]{0.33\linewidth}
\centering
\includegraphics[keepaspectratio, height=4cm]{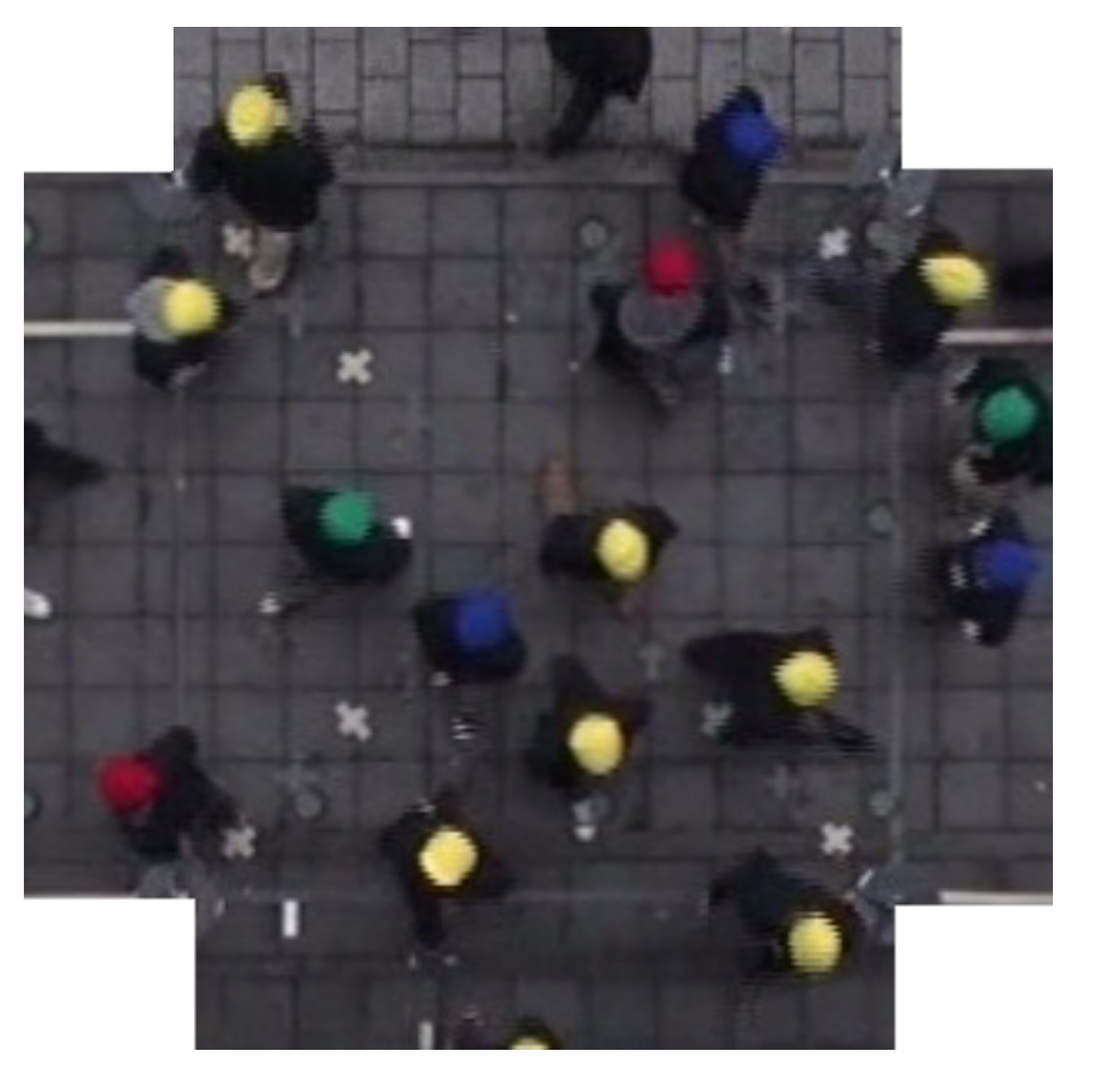}
\subcaption{}\label{fig:W3_0m_example}
\end{minipage}
\begin{minipage}[t]{0.33\linewidth}
\centering
\includegraphics[keepaspectratio, height=4cm]{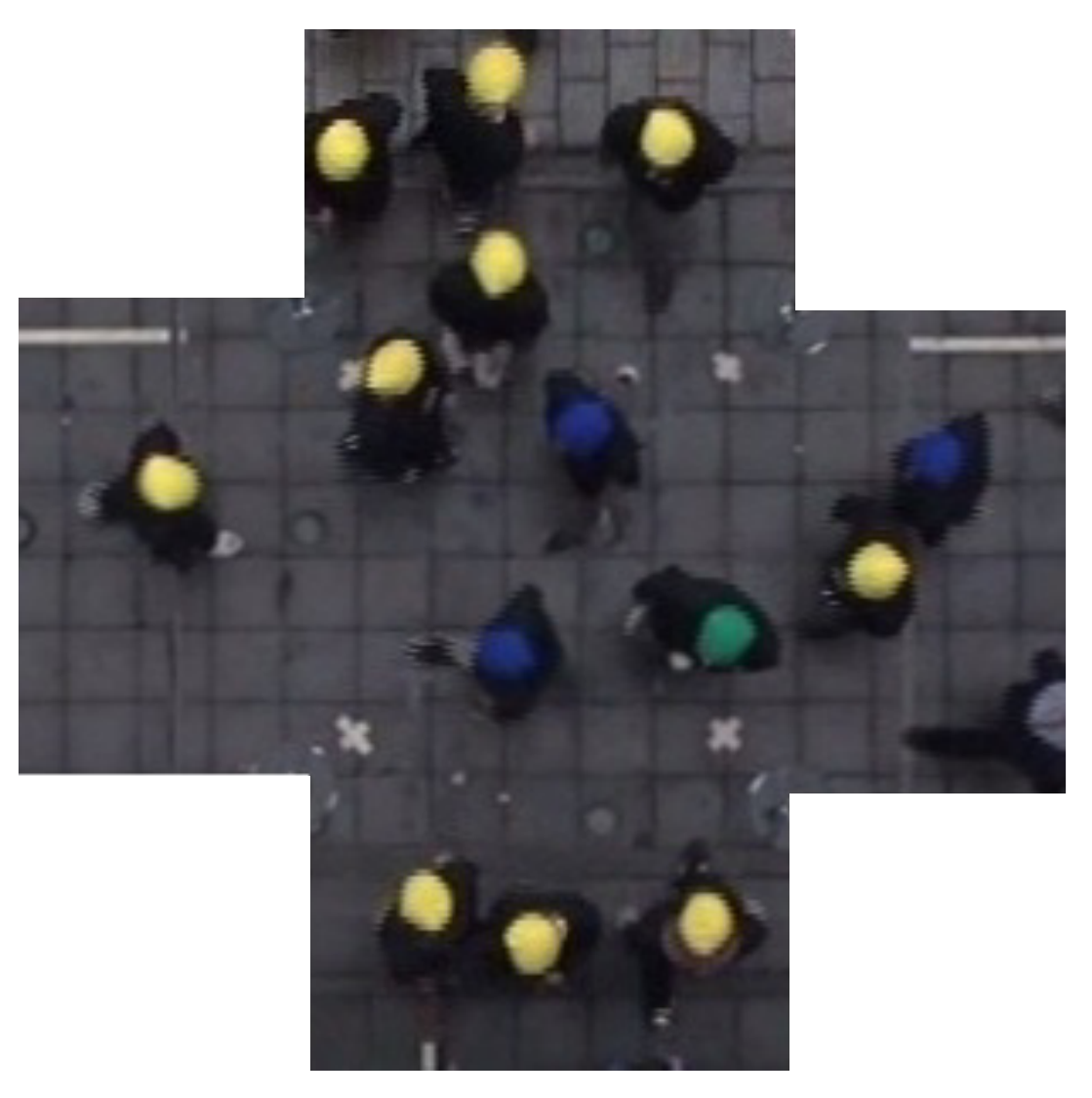}
\subcaption{}\label{fig:W2_0m_example}
\end{minipage}
\begin{minipage}[t]{0.33\linewidth}
\centering
\includegraphics[keepaspectratio, height=4cm]{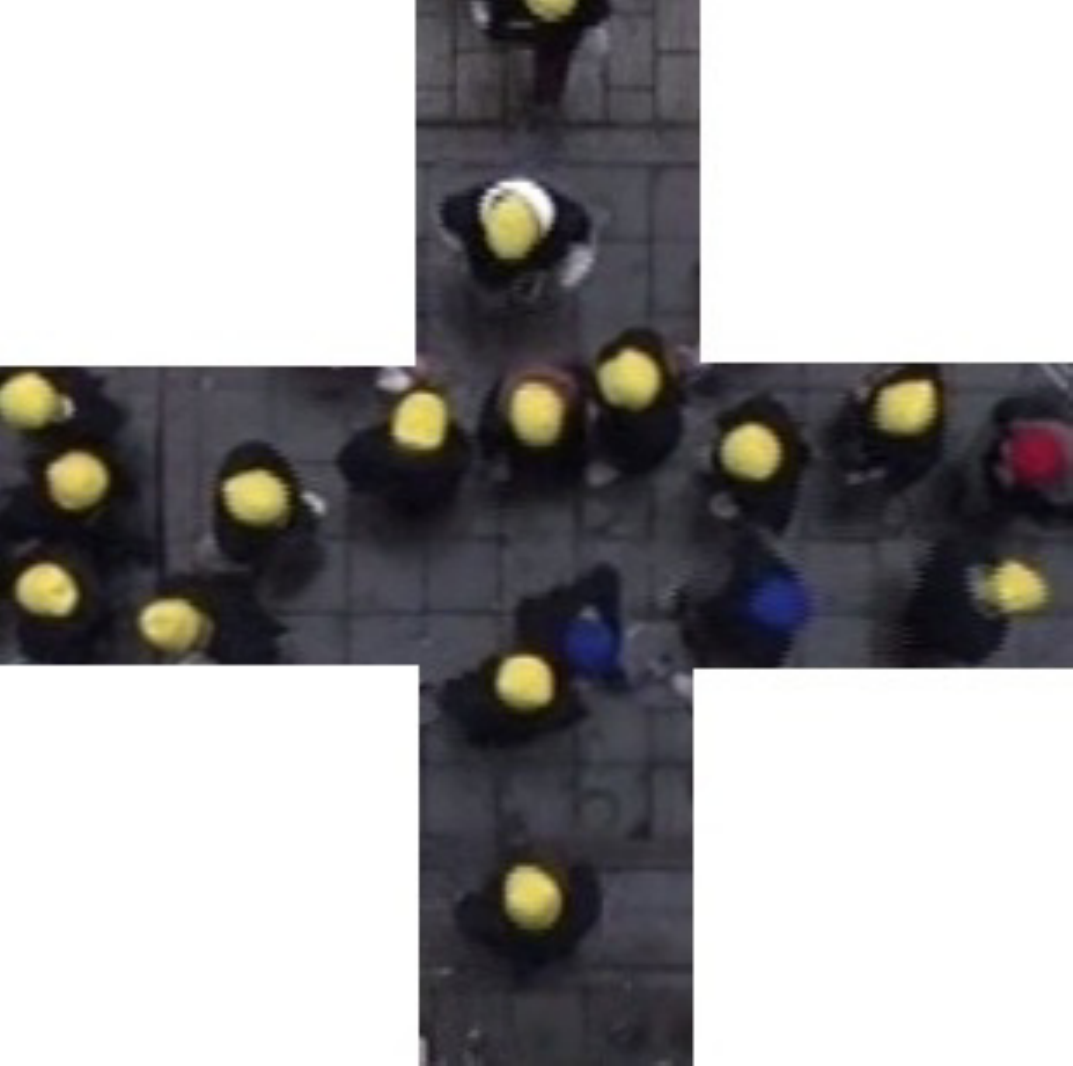}
\subcaption{}\label{fig:W1_0m_example}
\end{minipage}
\end{tabular}
\caption{
Video frames from the three different configurations.
(a) $W_{\rm{leg}}=3$ m, (b) $W_{\rm{leg}}=2$ m, and (c) $W_{\rm{leg}}=1$ m.
To focus on the experimental area,
inaccessible areas were cut from the images.
For each configuration, four executions were performed with the different initial conditions described in Tab.~\ref{tab:start_switch}.
}
\label{fig:video_frames}
\end{figure}

Finally, to check the impact of density on body rotation, three different widths ($W_{\rm{leg}}=3$, 2, and 1 m) were investigated for the intersecting legs.
For this purpose, the final parts of the four legs were restricted, creating a bottleneck before entering the intersection.
As shown in Fig.~\ref{fig:section_width}, a particular setup was chosen to realize a straight path before the intersection in each configuration.
Note that the starting positions (the lines of Fig.~\ref{fig:cross_geometry}) and experimental concepts described so far were unchanged.
Video frames relative to each configuration are shown in Fig.~\ref{fig:video_frames}.
Inaccessible areas created from changing the leg width 
were colored white to 
focus on the experimental area.

Overall, the presented experimental setup allowed for the recreation of a fairly natural cross flow in which participants had to consider each situation independently.
The careful planning concerning turning directions and starting positions enabled comparable situations, in which only density played an important role, to be considered.

Since it was difficult to distinguish turning and body rotation clearly for participants turning right and left at the intersection, the focus was placed on the participants equipped with the tablets and going straight, i.e., the participants 3 and 49 (Tab. \ref{tab:id_turning}) in the following analyses.
The analyses of the other participants are presented in \cite{Feliciani2017PhD}. 

\SUB{Discrepancy between body rotational angle and walking direction}{CrossRes1}


\FIGT{\EPSW{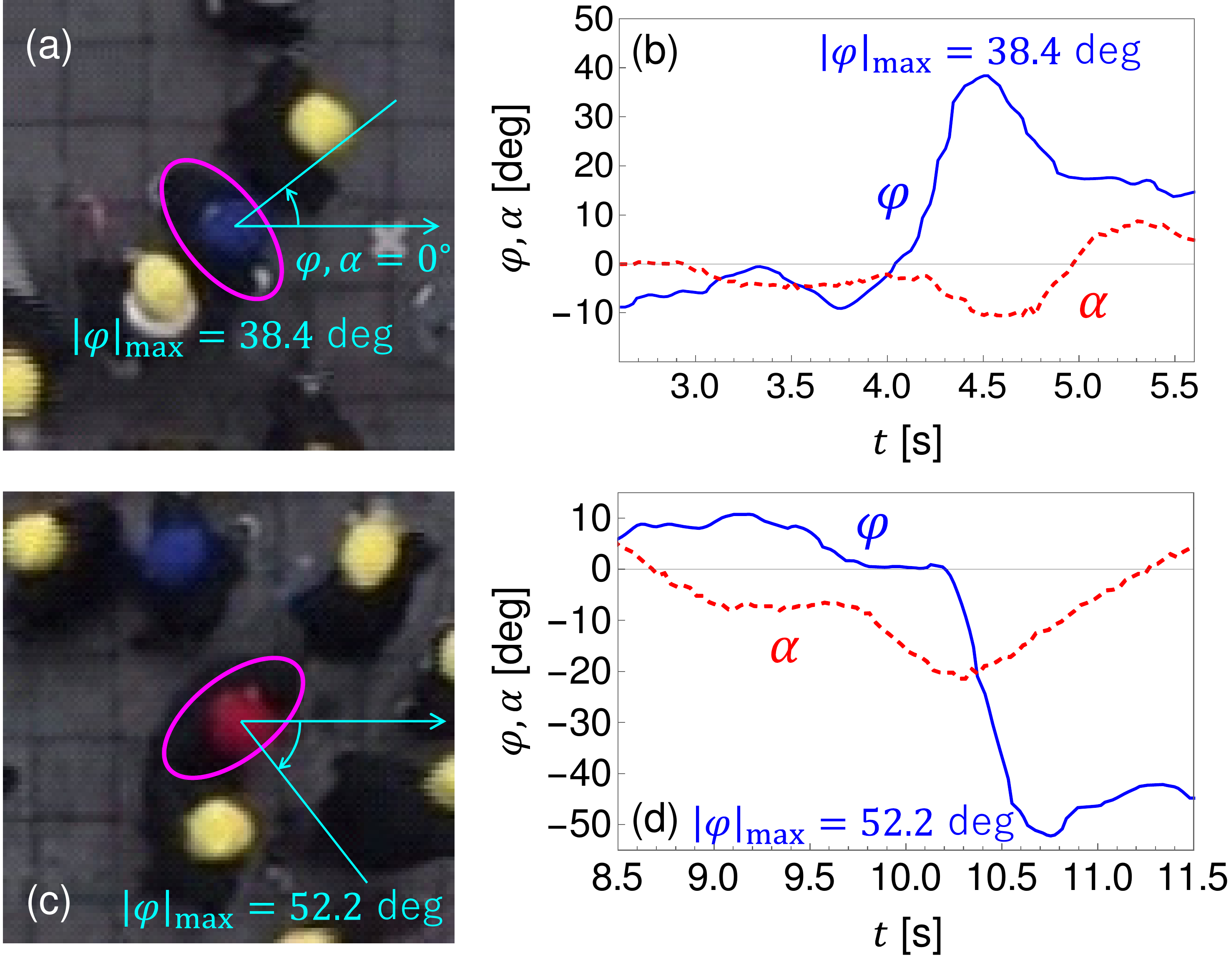}{16}{
Snapshots and evolution of body rotational angle $\varphi$ and walking direction $\alpha$ in the cross flow experiment for $W_{\rm{leg}}=2$ m.
The snapshots correspond to the times when the maximum absolute rotational angles $|\varphi|_{\rm{max}}$ were achieved.
(a), (b) The results of participant 3 with a blue cap walking from the West (Left) to the East (Right) in Execution 1.
(c), (d) The results of participant 49 with a red cap walking from the West (Left) to the East (Right) in Execution 3.
}}

Figure \ref{fig_cross_traj_gyro}a shows the snapshot of participant 3 with a blue cap walking from the West (Left) to the East (Right) in the case $W_{\rm{leg}}=2$ m.
It can be seen that he avoided other participants by rotating his body counter-clockwise and tried to move to the right.
This phenomenon is understood more clearly by analyzing the evolution of the body rotational angle $\varphi$ and walking direction $\alpha$, as in Fig. \ref{fig_cross_traj_gyro}b.
The body rotational angle $\varphi$ becomes large at $t=4.5$ s, and a great discrepancy was observed between the body rotational angle $\varphi$ and walking direction $\alpha$.
These results correspond well to the snapshot (Fig. \ref{fig_cross_traj_gyro}a), i.e., participant 3 successfully moved to the right without changing walking direction by using body rotation to avoid others.
This is an experimental fact that a pedestrian avoids others by rotating his body without greatly changing his walking direction.

Figures \ref{fig_cross_traj_gyro}c and \ref{fig_cross_traj_gyro}d show similar results.
Participant 49 with a red cap walked from the West (Left) to the East (Right) in the case $W_{\rm{leg}}=2$ m.
Here, he rotated his body clockwise and tried to walk through other participants.
Evolution in the walking direction indicates that participant 49 walked to the south-east direction with some degree, since the absolute walking direction achieved $|\alpha| \approx 20$ deg.
However, the absolute body rotational angle $|\varphi|$ achieved at $t=10.8$ s was much larger.
Therefore, participant 49 is considered to have rotated his body not only for changing his walking direction, but also for going through other participants.

The discrepancies between the body rotational angles (body orientation) and walking directions indicate that they must be dealt with separately.


\FIGT{\EPSW{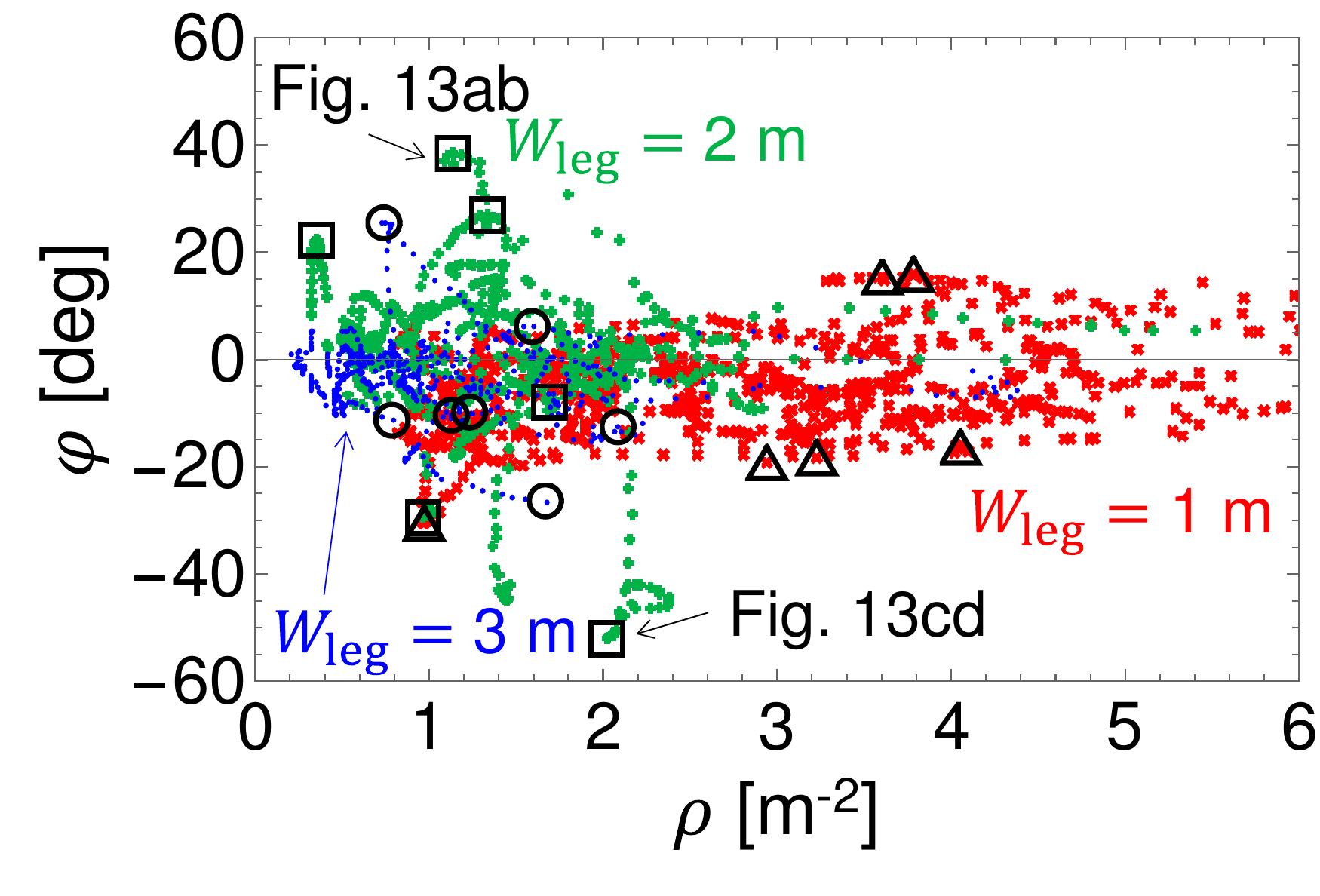}{10}{
Body rotational angle $\varphi$ as a function of the local density $\rho$ for participants 3 and 49.
The blue dots ($\cdot$), green crosses ($+$), and red crosses ($\times$) represent the instantaneous results for $W_{\rm{leg}}=3$, 2, and 1 m, respectively (29.97 frame/sec (NTSC)).
The black circles, squares, and triangles show the maximum absolute rotational angles in each trial for $W_{\rm{leg}}=3$, 2, and 1 m, respectively.
Note that the data, which correspond to the situations in Figs. \ref{fig_cross_traj_gyro}ab and \ref{fig_cross_traj_gyro}cd, are also plotted in the figure.
}}

\SUB{Relation between density and body rotational angle}{CrossRes2}
Next, the relationship between the density and body rotational angle were investigated.
Figure \ref{fig_cross_density_angle} shows the body rotational angle $\varphi$ as a function of the local density $\rho$ for participants 3 and 49.
The local densities $\rho$ were calculated by depicting the Voronoi diagram using the video camera images.
The blue, green, and red plots are the instantaneous results for $W_{\rm{leg}}=3$, 2, and 1 m, respectively.
Since the frame rate of the video camera (29.97 frame/sec (NTSC)) was lower than the frequency of the tablets (50 Hz), the resolution of the data was reduced to that of the video camera
\footnote{
Note that the successive data points are strongly correlated; however, this correlation does not harm our result. Since the time of one trial was long enough, we could observe various situations in one trial, in other words, the data are not biased toward specific situations.
If we decrease the sampling rate, the plots become sparse.
Furthermore, the maximum rotational angles may change because we sometimes fail to observe the larger rotational angles due to the low sampling rate.
Therefore, high sampling rate is important to see the body rotation by analyzing rotational angles.
However, we confirmed that the changes of the maximum rotational angles were small, i.e, the positions of the black circles, squares, and triangles in Fig. \ref{fig_cross_density_angle} do not greatly change by the sampling rate, due to the strong correlations between the successive data.
}.
Note that the data from the tablets were interpolated to depict the figure.
The black circles, squares, and triangles show the maximum absolute rotational angles in each trial for $W_{\rm{leg}}=3$, 2, and 1 m, respectively.

The body rotational angles were observed to achieve $\varphi \approx 20$ deg for a broad range of densities.
In the cross flow, the participants needed to avoid others from multiple directions, thus, they rotated their body to some degree even at low densities.
Furthermore, large body rotational angles were achieved in only the low and middle density regions ($\rho < 2.5$ m$^{-2}$).
When the density was not high, a participant could look for a space between the others and went through there by rotating his body.
It was confirmed that the large body rotational angles $\varphi$ observed for $\rho < 2.5$ m$^{-2}$ corresponded to such a phenomenon.
However, for higher densities ($\rho > 2.5$ m$^{-2}$), it was difficult for the participants to find enough space to go through.
Thus, they slowed down or stopped and waited until enough space appeared in front of them.
Hence, no large body rotational angles $\varphi$ could be seen in the high-density region.

As Fig. \ref{figure2} shows, the decrease in the width of an ellipse is small for small rotational angles.
Therefore, the conventional models can simulate cross flow well to some extent, even if they do not include body-rotation behavior, since most pedestrians do not rotate greatly.
However, such models cannot reproduce pedestrians going through others by body rotation, who were observed in our cross-flow experiment.
Hence, body-rotation behavior is indeed an indispensable factor for more realistic simulations.

\SEC{Conclusion}{CONC}
This study mainly investigated pedestrians' collision avoidance behaviors involving body rotation and side-stepping, when body orientation differs from walking direction, such as in a narrow corridor.
A pedestrian was modeled with an ellipse, and the mathematical relationship between the body rotational angles of two elliptic pedestrians and the corridor width was obtained.
Moreover, experiments with real pedestrians were conducted, and the rotational angles were measured using commercial tablets.
The results of the experiment indicate that pedestrians start to rotate when their shoulders may come into contact, and the rotational angles during passing increase as the corridor width decreases.
Finally, the sizes of the elliptic pedestrians were determined by the least squares method, using both the theoretical and experimental results.
The calibrated size of the elliptic pedestrians quantitatively agreed with the effective size of real pedestrians.
We would like to emphasize that the participants of our experiments included both males and females with various ages, so that we consider that our results are not limited to the specific types of pedestrians.
Furthermore, a simulation model for pedestrians passing through a narrow corridor was developed.
The simulation results agreed with the experimental results.
A cross-flow experiment was conducted to investigate body-rotation behavior in a more complex and realistic scenario.
The results indicated the importance of considering body-rotation behavior in complex scenarios.
It was also found that the participants tended to rotate greatly and stepped sideways when the density was not high.
When the density was high, the participants could not move forward even with physical rotations, and thus large body rotations were not observed.

In the future, the combination of position and rotational angle data will allow for the analysis and modeling of
when and how pedestrians start body rotation.
The coupling effect of changing walking direction and body orientation on collision avoidance will also be studied. 

\section*{Acknowledgments}
We would like to thank the entire team of the Nishinari group for their assistance during the planning and execution of the experiment.
We also appreciate Takahiro Ezaki for his useful comments regarding our manuscript.
The authors would like to thank Enago (www.enago.jp) for the English language review.
This work was supported by JSPS KAKENHI Grant Numbers 25287026 and 15K17583, and JST-Mirai Program Grant Number JPMJMI17D4, Japan.
 
\appendix

\SEC{Width of a rotated ellipse}{MATH}
Here, we provide details on derivation of the relationship between the passing rotational angle $\theta \in [0, \pi/2]$ and the width $d$ of an ellipse.
With proper choice of coordinate system, ellipse $i$ may be expressed quadratically as follows:
 \begin{equation}
\frac{x^2}{a^2}+\frac{y^2}{b^2}=1,
\label{eq:A1}
\end{equation}
where $a$ and $b$ are the semi-major and the semi-minor axes of the ellipse, that is, $a>b$. 

\begin{figure}[htbp]
\centering\hspace{1cm}\includegraphics[width=8cm,clip]{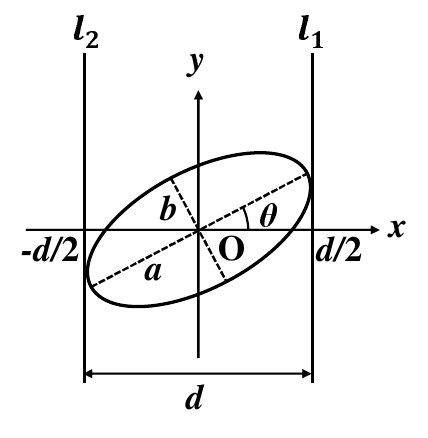}
\caption{
Schematic of rotated ellipse $i'$.
Geometric relationship between rotational angle $\theta$ and width $d$ is also shown in the figure.
}
\label{fig:ellipse}
\end{figure}

Next, we rotate ellipse $i$ counter-clockwise by $\theta$ with respect to its center $\rm{O}$ and consider ellipse $i'$, as in Fig. \ref{fig:ellipse}. 
Using rotational transformation, a point ($x_2, y_2$) on ellipse $i'$, which corresponds to a point ($x_1, y_1$) on ellipse $i$, is described as 
\begin{equation}
\left(
    \begin{array}{c}
x_2 \\
y_2
\end{array}
  \right)
=\left(
    \begin{array}{cc}
\cos\theta & -\sin\theta \\
\sin\theta & \cos\theta
\end{array}
  \right)
\left(
    \begin{array}{c}
x_1 \\
y_1
\end{array}
  \right).
\label{eq:A2}
\end{equation}
By deforming (\ref{eq:A2}), we obtain 
\begin{eqnarray}
\left\{
\begin{array}{l}
x_1=x_2\cos\theta+y_2\sin\theta, \\
y_1=-x_2\sin\theta+y_2\cos\theta.
\end{array}
\right.
\label{eq:A3}
\end{eqnarray}
Since the point ($x_1, y_1$) is on ellipse $i$, it satisfies (\ref{eq:A1}).
Thus, we substitute (\ref{eq:A3}) into (\ref{eq:A1}) and obtain the equation of the rotated ellipse $i'$ as follows:
 \begin{equation}
\frac{(x\cos\theta+y\sin\theta)^2}{a^2}+\frac{(-x\sin\theta+y\cos\theta)^2}{b^2}=1,
\label{eq:A4}
\end{equation}
where we replace $x_2$ and $y_2$ with $x$ and $y$, respectively.

We denote the tangents of ellipse $i'$ that are parallel to the $y$ axis as $l_1$ and $l_2$, as in Fig. \ref{fig:ellipse}.
Their equations are given as
\EQL{eq:A5}{\LP{
l_1: x = d/2, \\
l_2: x = - d/2,
}}
because ellipse $i'$ is symmetric with respect to its center $\rm{O}$.

Then, we consider the intersection equation of ellipse $i'$ and tangent $l_1$ ($l_2$) by substituting (\ref{eq:A5}) into (\ref{eq:A4}), which yields a quadratic equation with respect to $y$,
\begin{equation}
\frac{(d\cos\theta / 2 +y\sin\theta)^2}{a^2}+\frac{(-d\sin\theta / 2 +y\cos\theta)^2}{b^2}=1.
\label{eq:A6}
\end{equation}
Since $l_1$ ($l_2$) is tangential to ellipse $i'$, the discriminant of (\ref{eq:A6}) $D=0$.
By solving this equation, we finally obtain the relation between $\theta$ and $d$ as follows:
\begin{equation}
d(\theta) = 2 \sqrt{a^2\cos^2\theta+b^2\sin^2\theta}.
\label{eq:A8}
\end{equation}

Note that more complicated scenarios, for instance, consideration of the distance between ellipses and lines, have been analyzed in other studies \cite{Chraibi2010, Zheng2007, Chraibi2012rg}.

\SEC{Conversion method for passing rotational angle}{100cm}
In this appendix, we explain how we converted inappropriate passing rotational angles to appropriate ones.
We decided not to simply remove but include them for analysis by conversion because removing such data, whose values are small, puts improper focus on rare data with clear rotation and awkwardly increases the average passing rotational angles. 

As described in Sec. \ref{PRAngles}, we firstly determined the two thresholds:
one for the passing rotational angle $\theta_{i}$ and the other for the time gap between the points at which the participants achieved the maximum absolute rotational angle $\Delta t _{|\varphi|_{\rm max}}$.
The threshold for $\theta_{i}$ was calculated for each participant by averaging the maximal absolute rotational angles during passing
in $W=120$ and 140 cm.
In corridor experiment 1 (Secs. \ref{EXP} and \ref{COMP}), $\theta_{{\rm th}, 1} = 8.4^\circ$, $\theta_{{\rm th}, 2} = 8.5^\circ$, $\theta_{{\rm th}, 3} = 6.0^\circ$, and $\theta_{{\rm th}, 4} = 8.1^\circ$.
In corridor experiment 2 (Secs. \ref{EXP1-2} and \ref{COMP1-2}), $\theta_{{\rm th}, 0} = 10.8^\circ$, $\theta_{{\rm th}, 1} = 8.7^\circ$, $\theta_{{\rm th}, 2} = 9.7^\circ$, $\theta_{{\rm th}, 3} = 6.8^\circ$, $\theta_{{\rm th}, 4} = 7.5^\circ$, $\theta_{{\rm th}, 5} = 8.2^\circ$, $\theta_{{\rm th}, 6} = 10.1^\circ$, $\theta_{{\rm th}, 7} = 7.4^\circ$, and $\theta_{{\rm th}, 8} = 8.6^\circ$.
The threshold $\theta_{{\rm th}, i}$ represents the maximum rotational (oscillation) angle in the pass without clear rotation of participant $i$.
The threshold for $\Delta t _{|\varphi|_{\rm max}}$ was determined to be 0.5 s from Figs. \ref{fig:figure5} and \ref{figure6}.

Next, we applied the following procedure to all datasets corresponding to $W \le 100$ cm:
\EN{
\item For a datum with $|\varphi|_{{\rm max}, i} < \theta_{{\rm th}, i}$, we judged that there was no clear body rotation and used $\theta_{{\rm th}, i}$ as a passing rotational angle
\footnote{
As shown in Figs. \ref{fig:figure5}g and \ref{fig:figure5}h, we observed some oscillations of the rotational angles even though 
there were no clear body rotations.
Therefore, if we used 0$^\circ$ as the passing rotational angles in the cases of no clear body rotation, downward bias would arise in the average passing rotational angles.
Hence, $\theta_{{\rm th},i}$ was adopted as the passing rotational angle instead of 0$^\circ$.
}.
%
%
%
%
\item For a dataset with $\Delta t _{|\varphi|_{\rm max}} > 0.5$ s, $|\varphi|_{{\rm max}, i} > \theta_{{\rm th}, i}$ and  $|\varphi|_{{\rm max}, j} > \theta_{{\rm th}, j}$ $(|\varphi|_{{\rm max}, i} > |\varphi|_{{\rm max}, j})$, we judged that passing occurs at $t_{|\varphi|_{{\rm max}, i}}$, that is, the moment when $|\varphi|_{{\rm max}, i}$ was achieved.
Then, we determined passing rotational angles $\theta_i = |\varphi|_{{\rm max}, i}$ and $\theta_j$ as the larger of $\theta_{{\rm th}, j}$ and the maximum rotational angle achieved between $t_{|\varphi|_{{\rm max}, i}} - 0.5$ s and  $t_{|\varphi|_{{\rm max}, i}} + 0.5$ s.
}

In corridor experiment 1, step (i) was applied to 3 and 17 data for $W=90$ and 100 cm, respectively.
Step (ii) was applied to 2 and 7 data for $W=90$ and 100 cm, respectively.
In corridor experiment 2, step (i) was applied to 8 and 18 data for $W=90$ and 100 cm, respectively.
Step (ii) was applied to 10, 4, 6, 38, and 62 data for $W=60$, 70, 80, 90 and 100 cm, respectively.
Note that 48 and 144 data (24 and 72 datasets) were obtained for each corridor width $W$ in corridor experiment 1 and 2 as described in Secs. \ref{EXP} and \ref{EXP1-2}, respectively .

\section*{References}




\end{document}